\documentclass[10pt,aps,pra,twocolumn,superscriptaddress]{revtex4-2}

\usepackage[utf8]{inputenc}
\usepackage{enumerate}
\usepackage{braket}
\usepackage{amsmath}
\usepackage{amsthm}
\usepackage{amssymb}
\usepackage{amsfonts}
\usepackage{graphicx}
\theoremstyle{plain}
 
\usepackage{float}
\usepackage{bm}
\usepackage{hyperref}
\usepackage{fancyhdr}
\usepackage{xr}
\usepackage{xcite}
\usepackage{color}
\usepackage[dvipsnames]{xcolor}

\begin{document}

\title{Phase diagram of quantum generalized Potts-Hopfield neural networks}
\author{Eliana Fiorelli}
\thanks{e.fiorelli@fz-juelich.de}
\affiliation{Institute for Theoretical Nanoelectronics (PGI-2), Forschungszentrum J\"{u}lich, 52428 J\"{u}lich, Germany}
\affiliation{JARA-Institute for Quantum Information, RWTH Aachen University, 52056 Aachen, Germany} 
\author{Igor Lesanovsky}
\affiliation{School of Physics and Astronomy, University of Nottingham, Nottingham, NG7 2RD, UK}
\affiliation{Centre for the Mathematics and Theoretical Physics of Quantum Non-equilibrium Systems, University of Nottingham, Nottingham NG7 2RD, UK}
\affiliation{Institut für Theoretische Physik, Universität Tübingen, Auf der Morgenstelle 14, 72076 Tübingen, Germany}
\author{Markus M\"{u}ller}
\affiliation{Institute for Theoretical Nanoelectronics (PGI-2), Forschungszentrum J\"{u}lich, 52428 J\"{u}lich, Germany}
\affiliation{JARA-Institute for Quantum Information, RWTH Aachen University, 52056 Aachen, Germany}
\date{\today}

\begin{abstract} 
We introduce and analyze an open quantum generalization of the q-state Potts-Hopfield neural network, which is an associative memory model based on multi-level classical spins. The dynamics of this many-body system is formulated in terms of a Markovian master equation of Lindblad type, which allows to incorporate both probabilistic classical and coherent quantum processes on an equal footing. By employing a mean field description we investigate how classical fluctuations due to temperature and quantum fluctuations effectuated by coherent spin rotations affect the ability of the network to retrieve stored memory patterns. We construct the corresponding phase diagram, which in the low temperature regime displays pattern retrieval in analogy to the classical Potts-Hopfield neural network. When increasing quantum fluctuations, however, a limit cycle phase emerges, which has no classical counterpart. This shows that quantum effects can qualitatively alter the structure of the stationary state manifold with respect to the classical model, and potentially allow one to encode and retrieve novel types of patterns. 
\end{abstract}

\maketitle

\section{Introduction}
Machine Learning (ML) is considered the core of artificial intelligence and data science, and it is nowadays an expanding research area. The interest in giving systems the ability to learn how to accomplish a task, without being explicitly programmed, ranges from computer science to neurobiology \cite{Samuel59, GoodfellowEtAl16}, with applications spreading from science to commerce \cite{JordanM15, LeCunEtAl15}. One of the most successful architectures for ML is represented by neural networks (NNs) \cite{Haykin98}, which are artificial systems that are able to mimic what is known about the inner workings of the brain in relation to assimilation and comprehension processes. While in general these models may be too simple for the pursuit of understanding the computational properties of the brain itself \cite{Crick89}, they have nevertheless powerful applications in artificial intelligence. NNs are systems described in terms of interconnected artificial neurons with learning capabilities. Among the several instances of NNs, many efforts are focused on \textit{attractor} NNs, which are well suited to be described via classical spin systems subject to thermal fluctuations \cite{Amit_book}. This formulation enables one to adopt concepts commonly used in the context of statistical mechanics.

In the last decades, it has been shown that genuine quantum features such as superposition of states and entanglement provide remarkable advantages for solving many computational problems \cite{Montanaro16, NielsenC11, Shor:SIAM:1999, Grover:PRL:1997}, which has led to significant developments in the field of quantum computation. Within this research area, a number of current developments focus on exploring possible advantages in the design and realization of near-term quantum devices by applying concepts from classical NN computing to quantum systems \cite{CarleoEtAl19}. One example is given by suitably constructed classical NNs, which are used to approximate many-body quantum states, such as Restricted Boltzmann Machines \cite{CarleoT17, GaoD17, HuangM17}. Other recent works deal with more sophisticated architectures, such as deep NNs, for efficiently modelling quantum many-body states \cite{ChooEtAl18, Saito18, SharirEtAl20}, allowing to implement representations of mixed states, rather than pure states only \cite{TorlaiM18}. Also noteworthy in this regard are recent advances in relating the NNs description of many-body quantum states to tensor network states, with the intention to exploit the latter as a  tool for ML tasks \cite{GlasserEtAl18, LiuEtAl18, PastoriEtAl19, Clark18}.

In addition to the aforementioned paradigms, various efforts aim at harnessing the potential computational power of quantum generalizations of classical NNs \cite{Schuld:QInf:2014, DengEtAl17}. Here the goal is to first understand how to model such quantum versions, and subsequently to analyze the emergence of novel features embodied in their quantum nature \cite{Biamonte:Nat:2017}. Current instances of quantum formulations of NNs range from quantum algorithms and quantum circuit settings \cite{RebentrostEtAl18, AspuruC20, KilloranEtAl19, ManginiEtAl21, TorronteguiG19, KristensenEtAl21, CaoGG17, CongCL19, BeerEtAl20}, which include, e.g., the so-called feed-forward quantum NNs, to condensed matter systems \cite{PonsEtAl07, GopalakrishnanEtAl12, FiorelliEtAl20} and biological settings \cite{Behrman06, AkazawaEtAl00}. One approach for generalizing NNs into the quantum domain --- which we will also pursue here --- views a task-performing NN as a dynamical process that takes place within an ensemble of neurons that are represented by quantum systems \cite{Rotondo:JPA:2018, Fiorelli:PRA:2019, DiamantiniT06}. This perspective is inspired by Hopfield-type NNs \cite{Hopfield:1982}, which are, amongst attractor NNs, the simplest instances of associative memories, permitting the recognition, or retrieval, of patterns. Here, the retrieval of information is described by means of a classical non-equilibrium dynamics, where memories are stored as long-time solutions of a stochastic dynamics \cite{Amit_book, Amit:PRL:1985, AmitGS:1987}. Such a classical out-of-equilibrium dynamics can be generalized into the quantum domain via the formalism of open quantum systems \cite{Garrahan18}, which allows to introduce quantum generalizations of Hopfield-type NNs. This idea can be further generalized, as recently done in Ref. \cite{LewensteinEtAl20}, where quantum NNs are represented as dynamical maps. They evolve initial quantum states, which play the role of inputs of the quantum NN, and take them towards stored memories, which correspond to stationary states of the maps. 

\begin{figure}
\center
\includegraphics[width=\linewidth]{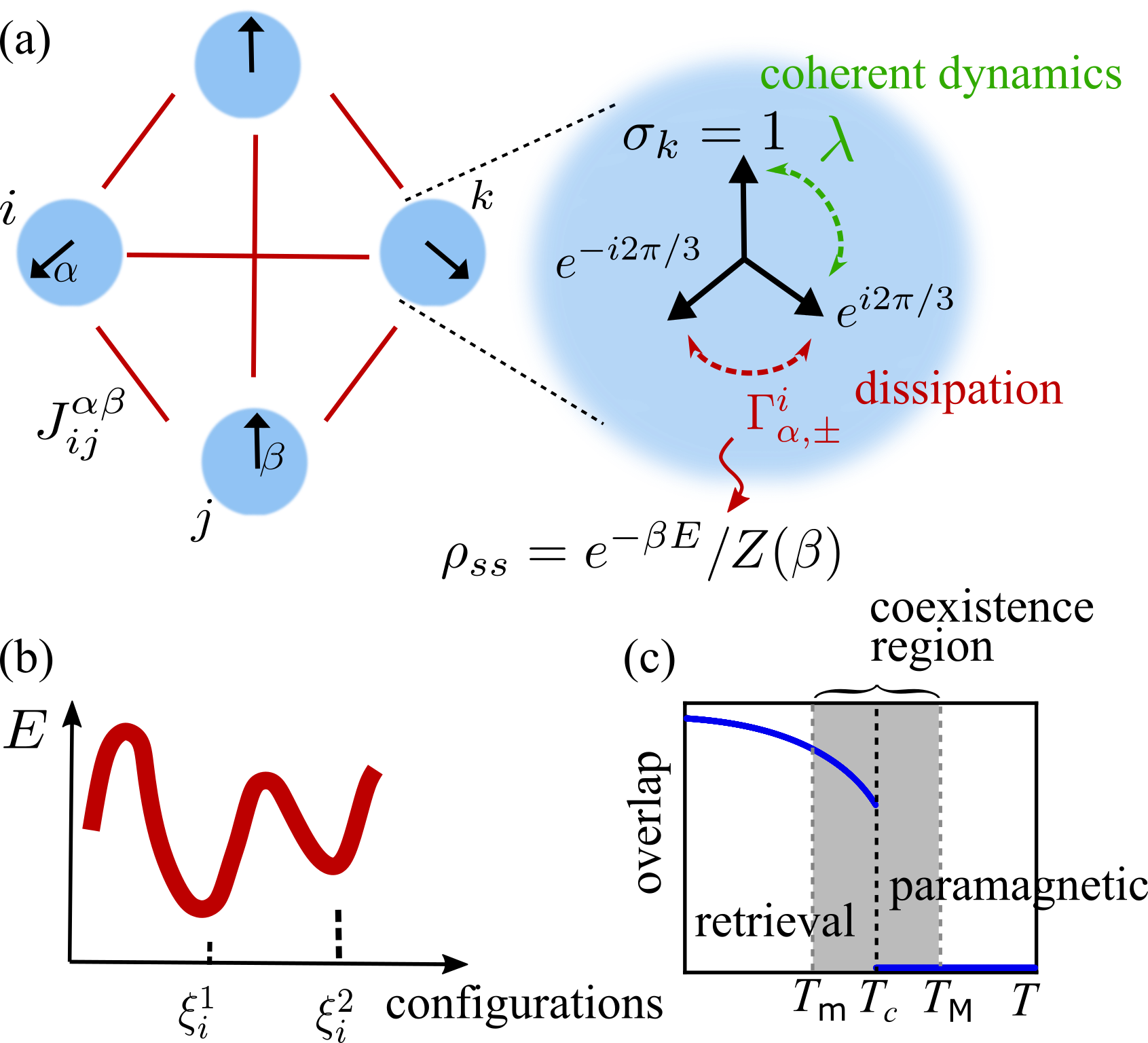}
\caption{\textbf{Quantum effects in the q-Potts-Hopfield neural network.} Scheme of the quantum generalization of the qPHN model. (a) All-to-all connected network of $N=4$ Potts spins. The $i$-th  and $j$-th spins (neurons) are in the state $\alpha$, $\beta$, respectively, and interact with coefficient $J_{ij}^{\alpha \beta}$. Here we set $q=3$, so that each neuron can assume the values $\lbrace 1, e^{i2\pi/3}, e^{-i2\pi/3} \rbrace$. A classical stochastic dynamics leading to an equilibrium state at temperature $T=1/\beta$ can be implemented via a purely dissipative Lindblad master equation. The thermal transitions between the internal states of a Potts spin occur with rate $\Gamma_{\alpha , \pm}^{i}$, which depends on the configuration of all the other neurons (see main text for details). Quantum effects are included by allowing coherent transitions (parametrized by the rate $\lambda$) between states of the Potts spins. (b) Sketch of the energy landscape of the fully qPHN. Memory states, i.e. patterns labeled as $\vec{\xi}^{\mu}$, are global minima of the energy function. In case of a purely dissipative dynamics, i.e. $\lambda = 0 $, the stationary state is a thermal one, $\rho_{ss}$, with respect to the energy function $E$ [see Eq.~\eqref{eq2_energyPotts1}]. When $\lambda \neq 0$, non-classical stationary states and also non-stationary long time behavior, such as limit cycle solutions, may emerge. (c) As a function of the temperature, the qPHN passes from a disordered, paramagnetic phase to an ordered one where patterns are retrieved. The transition occurs discontinuously at $T_c$, and it is signalled by the behavior of the overlap between Potts spin configurations and patterns [see Eq.~\eqref{e2_overlapCL}]. Within the temperature range $T_m \leq T \leq T_M$ a phase coexistence of paramagnetic and retrieval solutions takes place.} \label{fig_intro}
\end{figure}

In this work we introduce a quantum generalized NN, which is derived from the classical q-state Potts-Hopfield neural network (qPHN). Its basic constituents are multi-level spins which are subject to an all-to-all interaction, as shown in Fig.~\ref{fig_intro}(a). We introduce an open quantum dynamics that evolves this system by means of a Markovian dynamical map, using the Gorini-Kossakowski-Sudarshan-Lindblad (GKSL) quantum master equation \cite{BreuerP:2002}. This allows us to include the following competing effects: (i) dissipative, non-unitary processes represent the analogy with the classical out-of-equilibrium pattern-retrieval dynamics; (ii) coherent, unitary processes, which generate quantum fluctuations due to the creation of superpositions between spin states. In the context of quantum generalization of classical dynamical maps, the competition between coherent and dissipative dynamics has been explored in different settings, often providing the access to new classes of non-equilibrium many-body quantum systems and/or to novel phases of matter \cite{MullerEtAl12, SchindlerEtAl13}. Here, in order to investigate the impact of (ii) on the retrieval mechanism of the NN, we focus on analyzing the long-time state of the dynamics by means of a mean field approach. The derived phase diagram shows that the purely dissipative case, where coherent effects are absent, is indeed consistent with the classical pattern retrieval phenomenology. In the presence of quantum fluctuations the NN retains the capability to retrieve patterns. However, we find that due to the competition between dissipation and coherent processes a new limit cycle phase emerges. A similar phenomenon was previously observed within a quantum generalized Hopfield Neural Network (HNN) \cite{Rotondo:JPA:2018}, and we discuss similarities and differences with respect to the qPHN. Our study corroborates that quantum generalized Hopfield-type NNs feature novel phases which could potentially be exploited to define new types of retrieval. Moreover, they provide a setting for the realization of quantum maps with multiple stationary states, and therefore yield a physical platform for addressing questions concerning a possibly enhanced storage capacity of quantum NNs, as discussed in Ref.~\cite{LewensteinEtAl20}.

\section{q-state Potts-Hopfield neural network}
\label{Section 2}
In this section we review the main features of the classical model that describes a multi-level neural network referred to as q-state Potts-Hopfield neural network (qPHN) \cite{Kanter88, BolleDH_PA_92}. The basic constituents of this NN are $N$ multi-level classical spins, $\sigma_i$, $i = 1,...,N$, referred to as Potts spins. A single Potts spin can assume $q$ possible states which may be labeled by the roots of unity:
\begin{equation}\label{e2_PottsSpinCl}
\sigma_i=\omega^{k_i}, \; \omega= e^{i\frac{2\pi}{q}}, \; k_{i}=0,...,q-1,
\end{equation}
$\forall i = 1,...,N$. Information is stored in the form of special Potts spin configurations, or patterns, denoted by the vectors $\vec{\xi}^{\mu}=(\xi^{\mu}_{1},...,\xi^{\mu}_{N})$, with $\mu = 1,...,p$. Each component $\xi^{\mu}_{i}$ of a pattern can assume the values $\lbrace \omega^{k_i^{\mu}} \rbrace_{k_i^{\mu}=0,...,q-1}$. 
%
The interaction between Potts spins is governed by the energy functional
\begin{equation}\label{eq2_energyPotts1}
E=-\frac{1}{2}\sum_{i \neq j}^{N}\sum_{\alpha, \beta=1}^{q}J_{ij}^{\alpha \beta}u_{\sigma_i, \alpha } u_{\sigma_j, \beta},
\end{equation}
where $u_{\sigma_i, \sigma_j} \equiv q\delta_{\sigma_i, \sigma_j}-1$. The interaction strengths are parametrized by the symmetric connectivity matrix
\begin{equation}\label{eq2_Hebbs}
J_{ij}^{\alpha \beta}=\frac{1}{q^2 N}\sum_{\mu=1}^{p}u_{\xi_i^\mu,\alpha}u_{\xi_j^\mu, \beta}.
\end{equation}
A scheme of such an all-to-all network is given in Fig.~\ref{fig_intro}(a). With these definitions, and in the limit $p/N \ll 1$, i.e. a small number of patterns with respect to the network size, the patterns $\vec{\xi}^{\mu}$ are minima of the energy functional \eqref{eq2_energyPotts1}, as illustrated in Fig.~\ref{fig_intro}(b). The ratio of the number of patterns and Potts spins --- $\alpha = p/N$ --- is referred to as load parameter. When it increases above a critical threshold $\alpha_c$ spurious minima emerge, and the hence minima of the energy function can no longer be associated with stored patterns. The critical load paramater $\alpha_c$ thus yields the storage capacity, i.e. the maximum number of patterns that can be stored in a network of $N$ Potts spins.

For $\alpha < \alpha_c$, a pattern retrieval mechanism can be constructed via a classical discrete-time gradient descent dynamics with respect to the energy \eqref{eq2_energyPotts1}. Under this dynamics the $i$-th Potts spin, in a generic configuration $\sigma_i(t)$ at time $t$, is updated to a lower energy configuration, $\sigma'_{i}(t+1)$, at time $t+1$:
\begin{equation}
\sigma_i(t) \rightarrow \sigma'_i(t+1) \Leftrightarrow \Delta E_i = h_{\sigma'_{i}}- h_{\sigma_{i}} < 0.
\end{equation}
Here $\Delta E_i$ is the energy difference associated with the transition $\sigma_i \rightarrow \sigma'_i$, which depends on the potential energy of the $i$-th Potts spin in the state $\sigma_i$:
\begin{equation}\label{eq2_potentialPotts}
h_{\sigma_i} = -\sum_{j \neq i, j=1}^{N}\sum_{\alpha,\beta}J_{ij}^{\alpha \beta}u_{\sigma_i, \alpha}u_{\sigma_{j},\beta}.
\end{equation}
It is worth noting that $h_{\sigma_i}$ depends on both the states of all the other neurons, $j \neq i$, and the state of the $i$-th neuron itself. 

The pattern retrieval via gradient descent can be interpreted as a dynamics that takes the system of Potts spins to the equilibrium state with respect to the energy \eqref{eq2_energyPotts1} at zero temperature. Noise can be included in the NN by introducing a finite inverse temperature $\beta=1/T$ and a stochastic dynamics that leads the system to an equilibrium distribution $P_{eq}=e^{-\beta E}/Z(\beta)$, with $Z(\beta)$ denoting the partition function. Let $p(\vec{\sigma},t)$ be the probability to find the system in the configuration $\vec{\sigma}=(\sigma_1,...,\sigma_N)^{T}$ at time $t$. The stochastic dynamics with the stationary equilibrium state
$P_{eq}=\lim_{t\rightarrow\infty} p(\vec{\sigma},t)$ is then given by
\begin{equation}\label{eq_qPNN_stocastic}
\dot{p}(\vec{\sigma})=\sum_{i=1}^{N}\sum_{\sigma_i' \neq \sigma_i}w_{\sigma_i' \rightarrow \sigma_i}p(\vec{\sigma}_i')-w_{\sigma_i \rightarrow \sigma_i'}p(\vec{\sigma}_{i}).
\end{equation}
The transition rate for changing the state of the $i$-th Potts spin is
\begin{equation}\label{e2_ratesCl}
w_{\sigma_i \rightarrow \sigma_{i}'}=\frac{1}{Z_i}e^{-\beta \Delta E_i},
\end{equation}
with $Z_{i}=\sum_{\sigma_i=1}^{q}\sum_{\sigma' \neq \sigma}e^{-\beta(h_{\sigma'_i } - h_{\sigma_i})/2}$ \cite{BolleM_JPA_89}.

To analyze the pattern retrieval dynamics of the qPHN it is useful to introduce the overlaps between the Potts spins and the patterns. These play the role of order parameters, and their introduction drastically reduces the complexity of the problem (instead of the $q_N$-dimensional probability vector one considers an order parameter with $p$ components). The overlap of the spin configuration with the $\mu$-th pattern is given by
\begin{equation}\label{e2_overlapCL}
m^{\mu}=\frac{1}{N(q-1)}\sum_{i=1}^{N}u_{\xi_i^{\mu},\sigma_i}.
\end{equation}
Positive values of the overlap correspond to the retrieval of patterns, which is maximum when the overlap assumes its largest value. For instance, $m^{\mu}=1$ signals the complete retrieval of the $\mu$-th pattern. On the contrary, negative values of the overlap do not correspond to the storage of any of the patterns \cite{BolleEtAl_JPA_91, BolleM_JPA_89}. For sufficiently low temperatures, i.e. below a temperature $T_c$, thermal fluctuations do not affect the ability of the network to retrieve patterns. Here, positive values of the overlap are stable stationary solutions, allowing the retrieval of one pattern, when the corresponding overlap, say $m^{\mu}$, assumes a positive value in the stationary state. This characterizes the retrieval phase of the network. Negative solutions of the overlap can also appear as stable stationary solutions. However, for the cases that have been analysed, they disappear first than the retrieval solutions as the temperature is increased \cite{BolleM_JPA_89}. For $T > T_{c}$ thermal fluctuations prevent patterns to be stored as long time solutions, and the stationary state is characterized by vanishing overlaps, $m^{\mu}=0$, identifying a paramagnetic phase. At the critical temperature $T_c$ a first order phase transition between the retrieval and the paramagnetic phase takes place \cite{BolleDH_PRA_92}, where the overlap changes discontinuously from $ m^{\mu} > 0$ to $m^{\mu}=0$. Moreover, for temperatures within the interval $T \in [T_m,T_M] $, with $T_m < T_c < T_M$, both retrieval solutions and the paramagnetic one are present, this giving rise to a coexistence region. Here, retrieval (paramagnetic) solutions are stable (unstable) for $T \in [T_m,T_c]$, and unstable (stable) for $T \in [T_c,T_M]$, as highlighted in Fig.~\ref{fig_intro}(c). In the case of a qPHN with $q=3$, it is $T_m=2$, $T_c=2.14$, $T_M=2.18$ \cite{BolleEtAl_JPA_91}, with negative overlap solutions disappearing at $T=T_m$ \cite{BolleM_JPA_89}, and becoming unstable at $T=0$ for $p>1$ \cite{BolleEtAl_JPA_91}.

To conclude this section, let us focus on the particularly simple case of a qPHN storing one pattern, i.e. $p=1$. Here, one can apply the gauge transformation $\sigma_i\rightarrow \xi_i^{1}\sigma_i$, which aligns all the components of the pattern. This situation corresponds to the standard Potts model \cite{Wu82} described by the Hamiltonian $H=-\sum_{i,j=1}^{N}J_{i,j}u_{\sigma_i, \sigma_j} $. The latter further reduces to the all-to-all Ising model for the choice $q=2$. For a generic number $p$ of memories, the two-state Potts-Hopfield NN, $q=2$, corresponds to the Hopfield Neural Network (HNN) \cite{Hopfield:1982}, $E_H=-\sum_{i,j=1}^{N}J_{i,j}\sigma_i, \sigma_j $, with $\sigma_i = \lbrace \pm 1\rbrace$ and $J_{ij}$ derived from Eq.~\eqref{eq2_Hebbs} by setting $q=2$. It is worth noticing that, despite the fact that the HNN can be derived from the qPHN, and that it features the associative memory behavior, some differences can be highlighted. Firstly, the energy function of the HNN is characterized by a $\mathbb{Z}_2$-symmetry. As a consequence, given a pattern $\lbrace \xi^{\mu}_{1}, ..., \xi_{N}^{\mu}\rbrace$, its opposite one, $\lbrace -\xi^{\mu}_{1}, ..., -\xi_{N}^{\mu} \rbrace$, is also a stored memory, corresponding to negative value of the overlap. Contrarily, this symmetry is absent in the qPHN case for $q \geq 3$, and indeed negative values of the overlap do not correspond to storage of patterns. Secondly, when the thermal fluctuations are considered in the HNN, it undergoes a second order phase transition form a paramagnetic to a retrieval phase when the temperature is lowered, instead of a first order transition that is characteristic for the qPHN.

In the next sections, we will introduce a quantum generalization of the qPHN. The type of patterns we will focus on in this work are $(i)$ i.i.d. random variables, with probability distribution $\mathbf{P}(\xi_i^{\mu}=\omega^{k_i^{\mu}})=1/q$; $(ii)$ almost orthogonal patterns, satisfying $\lim_{N \rightarrow + \infty } \vec{\xi}^{\mu}\cdot  \vec{\xi}^{\nu} = \delta_{\mu,\nu} $, so that they can be considered distinguishable; furthermore we will focus on the limit of zero load, $p/N \ll 1$.

\section{Quantum model}

\label{Section 3}
In order to include  quantum effects in the qPHN, hereafter referred to as "quantum qPHN", we first promote the classical variables $\sigma_i$ to quantum operators $\hat{\Omega}_i$. Their eigenstates are $\ket{k_i}$, with $k_i=0,...,q-1$, and the eigenvalue equation is $\hat{\Omega}_i\ket{k_i} = \omega^{k_i}\ket{k_i}$. Transitions between states of a q-Potts spin are effectuated by the operators $\hat{T}^{\pm}_i$, such that $\hat{T}^{\pm}_i\ket{k_i}=\ket{k_i \pm 1}$, with $\hat{T}_i^{+}\ket{(q-1)_i}=\ket{0_i}$ and $\hat{T}_i^{-}\ket{0_i}=\ket{(q-1)_i}$. It is worth noting that the Potts spin operators obey the $\mathbb{Z}_q$-algebra
\begin{eqnarray}\label{eq_Zqalgebra1}
& \hat{T}_i^{+}\hat{\Omega}_i=\omega\hat{\Omega}_i\hat{T}_i^{+},\\\label{eq_Zqalgebra2}
& \hat{T}_i^{-}\hat{\Omega}_i=\omega^{*}\hat{\Omega}_i\hat{T}_i^{-}, \\\label{eq_Zqalgebra3}
& (\hat{T}_i^{+})^{q}=(\hat{T}_i^{-})^{q}=\hat{\Omega}_i^q= \mathbf{1}_{q}.
\end{eqnarray}

The first step towards including quantum effects into the qPHN is to formulate the classical stochastic dynamics of Eq.~\eqref{eq_qPNN_stocastic} in terms of a purely dissipative Markovian evolution, as given by the GKSL equation
\begin{equation}\label{eq_master_equation}
\dot{\rho} = \mathcal{L}(\rho)=\sum_{l}\hat{L}_l \rho \hat{L}^{\dagger}_l - \frac{1}{2} \lbrace  \hat{L}^{\dagger}_l  \hat{L}_l , \rho \rbrace.
\end{equation}
Here, $\rho$ is the state (density operator) of $N$ quantum q-Potts spins, and $\hat{L}_l$ are the jump operators. The latter are chosen such that they give rise to the same local processes described by the aforementioned classical dynamics. This means that the diagonal of the density operator $\rho$ evolves under exactly the same dynamics as the probability distribution of the classical qPHN. The jump operators that achieve this are
\begin{equation}\label{eq_newJumps}
\begin{split}
& \hat{L}_{\alpha,s}^{i}=\sqrt{\gamma}~\hat{\Gamma}_{\alpha,s}^{i}\hat{T}_{\alpha,s}^{i}, \\
& \hat{\Gamma}_{\alpha,s}^{i}=\frac{1}{\sqrt{\hat{Z}_i}}e^{\frac{\beta}{2}\Delta\hat{E}_{\alpha,s}^i},
\end{split}
\end{equation}
for $i=1,...,N$, $ \alpha=1,...,q$, $s=\pm 1$. Here the operator $\hat{T}_{\alpha, s}^{i}$ implements the transition of the $i$-th Potts spin from the state $\ket{\alpha}$ to the state $\ket{\alpha + s}$, and the operator $ \hat{\Gamma}_{\alpha,s}^{i}$ represents the operatorial form of the classical transition rate defined by Eq.~\eqref{e2_ratesCl}. The expression in the exponent reads
\begin{equation}\label{eq_newEnergy}
\Delta \hat{E}_{\alpha,s}^i=-\frac{1}{2}\sum_{j \neq i}^{N} \sum_{\eta, \eta'=1}^{q-1} \left[ \mathcal{J}^{\eta, \eta'}_{ij}\hat{\Omega}^{\eta'}_{j}\omega^{(\alpha-1) \eta}(\omega^{s \eta}-1)+\mathrm{h.c.} \right],
\end{equation}
with $\mathcal{J}_{ij}^{\eta \eta'}=\frac{1}{N}\sum_{\mu=1}^{p}(\xi_{i}^{*,\mu})^{\eta}(\xi_{j}^{*,\mu})^{\eta'}$,  and can be regarded as the corresponding change of energy under such transition. Analogously, the operatorial formulation of the classical partition function reads $\hat{Z}_i =  \sum_{\alpha} \sum_{s} e^{-\beta \Delta \hat{E}_{\alpha, s}^{i}} $. With these definitions, we can see that the operators $\hat{\Gamma}_{\alpha, s}^{i}$ are non-local ones, as they depend on the Potts-operators $\hat{\Omega}_j$ $\forall j \neq i$, i.e. the transition of the $i$-th Potts-spin from the state $\ket{\alpha}$ to the state $\ket{\alpha + s}$ is ruled by a rate which depends on the state of all the rest of the network.
The expression defined by Eq.~\eqref{eq_newEnergy} can be understood as the energy difference related to the operator
\begin{equation}
\hat{E}= -\frac{1}{2N}\sum_{\mu=1}^{p}\left\lbrace \left[\sum_{i=1 }^{N}\sum_{\alpha=1}^{q-1}(\xi_i^{*,\mu}\hat{\Omega}_i)^{\alpha} \right]^2 + \mathrm{h.c.} \right\rbrace,
\end{equation}
which can be obtained from the classical Potts-Hopfield energy, defined by Eq.~\eqref{eq2_energyPotts1}, by substituting classical Potts-spin variables with their quantum operators, furthermore requiring to satisfy hermiticity.

Using the master equation formulation of the qPHN dynamics immediately allows the inclusion of quantum effects through a Hamiltonian. More specifically, we consider the latter being a non-commuting term with respect to the set of jump operators \eqref{eq_newJumps}. This leads to the population of off-diagonal terms in the density operator $\rho$, when written with respect to the classical basis $\lbrace \otimes_{i=1}^{N} \ket{k_{i}} \rbrace$. For the prupose of the present work, we choose a simple Hamiltonian \cite{RotondoEtal:2018}, which reads 
\begin{equation}\label{eq_Hterm}
\hat{H}=\lambda \sum_{i=1}^{N}\hat{T}_{i},
\end{equation}
with $\hat{T}^i=\hat{T}_{+}^i+\hat{T}_{-}^{i}$. Given the non-commutativity of such a term and the rate components $\hat{\Gamma}_{\alpha, s}^{i}$ of the jump operators \eqref{eq_Hterm}, it acts on the Potts spins as a "transverse field", i.e. it will create superpositions between classical basis states. This leads to the formation of coherences and concomitant quantum fluctuations. Combining the adjoint Eq.~\eqref{eq_master_equation} with the adjoint von Neumann equation of the Hamiltonian \eqref{eq_Hterm} yields the equation of motion (EoM) of a generic operator $\hat{O}$:
\begin{equation}\label{eq_Lindblad_op}
\dot{\hat{O}}=i[\hat{H},\hat{O}]+\sum_{l}\hat{L}_{l}^{ \dagger} \hat{O} \hat{L}_{l}- \frac{1}{2}\lbrace \hat{L}_{l}^{ \dagger}\hat{L}_{l},  \hat{O} \rbrace.
\end{equation}

Equipped with the operatorial evolution provided by Eq.~\eqref{eq_Lindblad_op}, we can analyze how the competition between the classical qPHN dynamics and the Hamiltonian dynamics modifies the retrieval mechanism of the network. As such, we consider the functional form of the classical order parameter \eqref{e2_overlapCL} and promote it to be a quantum operator, getting
\begin{equation}\label{eq2_PottsNNoverlap}
\hat{m}^{\mu}= \frac{1}{2N(q-1)}\sum_{i=1}^{N}\sum_{\alpha=1}^{q-1}(\xi_{i}^{*,\mu}\hat{\Omega}_i)^{\alpha}+\mathrm{h.c.}
\end{equation}
We will thus investigate the expectation value of such a quantity, $\braket{\hat{m}^{\mu}}$, that can thus be interpreted as the overlap between the pattern configurations and the Potts-spin one. To do so, we will consider the stationary solutions of EoMs of the type of Eq.~\eqref{eq_Lindblad_op}, at varying of the coherent control parameter $\lambda$ and the temperature $T$, and setting $\gamma = 1$ in the definition Eq.~\eqref{eq_newJumps} of the jump operators.

\section{q=3 quantum Potts Neural Network}
We now specialize the generic quantum qPHN described in Sec.~\ref{Section 3} by fixing $q$, the number of levels for each Potts spin state. In this respect, we consider the simplest model of the quantum qPHN that goes beyond the $q=2$ case, which is equivalent to a HNN and has been explored in \cite{RotondoEtal:2018}. Thus, we will set $ q=3$, focusing on this case throughout the remainder of this work. In this section, we explain the main steps for deriving a set of EoMs that allow us to study the retrieval properties of the $q=3$ quantum qPHN, leaving a more detailed description of the derivation in Appendix \ref{Appendix B}.

\subsection{Collective operators}
We aim at obtaining a closed set of EoMs for the $N$ Potts spins, by employing the smallest possible number of operators. Such a set is given by the Potts spin operators $\hat{\Omega}_i$ and $\hat{\Omega}_{i}^{\dagger}$, previously defined in Sec.~\ref{Section 3}, and the following additional ones
\begin{equation}
\begin{split}\label{eq_X_Y}
& \hat{Y}_{\alpha}^{i}= \frac{1}{i}[\hat{T}_{\alpha,+}^{i}-(\hat{T}_{\alpha,+}^{i})^{\dagger}], \\
& \hat{X}_{\alpha}^{i}= \hat{T}_{\alpha,-}^{i}+(\hat{T}_{\alpha,-}^{i})^{\dagger}.
\end{split}
\end{equation}
Referring to the EoM \eqref{eq_master_equation} for the density operator, written in the classical basis $\lbrace \otimes_{i=1}^{N} \ket{k_{i}} \rbrace$, the dynamics of the operators $\hat{\Omega}_{i}$, $\hat{\Omega}_{i}^{\dagger}$ corresponds to the evolution of the populations, whereas the operators $\hat{X}_{\alpha}^{i}$ and $\hat{Y}_{\alpha}^{i}$ capture the off-diagonal elements, i.e.~the coherences.
The EoM for the operator $\hat{\Omega}_{i}$ reads
\begin{equation}\label{eq2_omega_coherent}
\dot{\hat{\Omega}}_{i} = -\frac{\gamma}{2}\hat{\Omega}_{i} -\gamma[\hat{f}_{1}^{i}\hat{\Omega}_{i}+\hat{f}_{2}^{i}\hat{\Omega}_{i}^{\dagger}+\hat{f}_{3}^{i}]-\lambda \sum_{\alpha=1}^{3}\omega^{\alpha-1}(\omega-1)\hat{Y}_{\alpha}^{i},
\end{equation}
where $\hat{f}^{i}_{\alpha}$, $i=1,...,N$ and $\alpha = 1,2,3$ are given as combinations of the operators $\hat{\Gamma}^{i}_{\alpha, s}$ and defined by Eqs.~\eqref{eq_appB_f1}-\eqref{eq_appB_f3}. The EoMs for the operators $\hat{X}_{\alpha}^{i}$ and $\hat{Y}_{\alpha}^{i}$ are more involved, given that such operators do not commute with the rates $\hat{\Gamma}^{j}_{\alpha', s}$, and crossing terms with $j \neq i$ are in principle present in the corresponding Lindblad equation. However, such additional terms are expected to scale as $O(p/N)$ \cite{Rotondo:JPA:2018}, and can thus be neglected in the thermodynamic limit (see also Appendix \ref{Appendix B}) that we enforce in the following. This yields
\begin{equation}\label{eq_eomX_Y}
\begin{split}
& \dot{\hat{X}}_{\alpha}^{i}=-\frac{\gamma}{2}h_{i}^{-}(\alpha)\hat{X}_{\alpha}^{i}-\lambda( \hat{Y}_{\alpha-2}^{i}- \hat{Y}_{\alpha}^{i}), \\
&  \dot{\hat{Y}}_{\alpha}^{i}=-\frac{\gamma}{2}h_{i}^{+}(\alpha)\hat{Y}_{\alpha}^{i}+ \lambda [\hat{X}_{\alpha +2}^{i}-\hat{X}_{\alpha}^{i}+2(\hat{P}_{\alpha+1}-\hat{P}_{\alpha})],
\end{split}
\end{equation}
where $h^{s}_{i}(\alpha)=\sum_{s'=\pm}(\Gamma_{\alpha-s',s'}^{i,2}+\Gamma_{\alpha-s'+s,s'}^{i,2})$. The operators $\hat{P}_{\eta}^i$ are projectors on the classical basis $\lbrace \ket{(\eta-1)_{i}} \rbrace$ of the $i$-th Potts spin, and are given by Eq.~\eqref{Ab_projectors} in terms of the operators $\hat{\Omega}_{i}$, $\hat{\Omega}_{i}^{\dagger}$.
\subsection{Equations of motion}
We are interested in investigating the behavior of the network by analyzing the expectation values of the overlap operator \eqref{eq2_PottsNNoverlap}. The latter will allow us to identify the presence or the absence of pattern retrieval. In the considered case, $q=3$, this is given by 
\begin{eqnarray}
\hat{m}^{\mu}=\frac{1}{2N} \sum_{i=1}^{N}(\xi^{*,\mu}_{i}\hat{\Omega}_{i}+ \xi^{\mu}_{i}\hat{\Omega}_{i}^{\dagger}).
\end{eqnarray}
In the presence of quantum coherences we have to consider four further collective observables, which dynamically couple to $\hat{m}^{\mu}$. These are constructed from the single spin operators \eqref{eq_X_Y}: 
\begin{equation}\label{eq_x_y_collective}
\begin{split}
& 
\hat{x}^{\mu}=\frac{1}{6N}\sum_{\alpha=1}^{3}  \sum_{i=1}^{N} \omega^{\alpha}\xi_{i}^{\mu} \hat{X}_{\alpha}^{i} + \mathrm{h. c. }, \\
&  \hat{\bar{x}}^{\mu}=i \left[\frac{1}{6N}\sum_{\alpha=1}^{3}\sum_{i=1}^{N}\omega^{-\alpha}\xi_{i}^{*,\mu} \hat{X}_{\alpha}^{i} - \mathrm{h. c. } \right], \\
& \hat{y}^{\mu}=\frac{1}{6N}\sum_{\alpha=1}^{3}\sum_{i=1}^{N}\omega^{\alpha} \xi_{i}^{\mu} \hat{Y}_{\alpha}^{i}  + \mathrm{h. c. }, \\ 
& \hat{\bar{y}}^{\mu}=i \left[\frac{1}{6N}\sum_{\alpha=1}^{3} \sum_{i=1}^{N} \omega^{-\alpha} \xi_{i}^{*,\mu} \hat{Y}_{\alpha}^{i}  - \mathrm{h. c. }\right].
\end{split}
\end{equation}
These collective operators are constructed such that their EoMs, together with that of $\hat{m}^{\mu}$, form a closed set, provided that the following approximations are performed (see Appendix \ref{Appendix B} for details): $(i)$ We neglect correlations amongst operators, which amounts to a mean field approximation. This means that, given a collective operator $\hat{O}^{\mu}$, we derive the EoMs for the quantity $O^{\mu} \equiv \braket{\hat{O}^{\mu}} $, using $\braket{\hat{O}^{\mu}\hat{O}^{\nu}} \approx \braket{\hat{O}^{\mu}}\braket{\hat{O}^{\nu}}$. Given the all-to-all coupling of the Potts spin model this approximation is expected to yield exact results in the thermodynamic limit \cite{Gayrard92}. $(ii)$ Since the stationary state is built upon the classical stochastic dynamics, we expect the dynamics toward stationarity to be mostly determined by the properties of the overlaps $m^{\mu}$, and therefore we replace $\braket{\hat{\Omega}_{i}} \approx \xi_{\mu}^{i}m^{\mu}.$ $(iii)$ We assume that the system exhibits the property of self averaging, i.e.~that, given a generic function $g(\xi_i^{\mu})$ of the patterns, for large $N$ we can perform the substitution $$\frac{1}{N}\sum_{i}g(\xi_{i}^{\mu}) \rightarrow \sum_{\lbrace \xi^{\mu} \rbrace} \mathbf{P}(\xi)g(\xi^{\mu}) \equiv \braket{\braket{g(\xi^{\mu})}},$$ where $\mathbf{P}(\xi)$ identifies the pattern probability distribution, chosen such that $\mathbf{P}(\xi_i^{\mu}=\omega^{k_i^{\mu}})=1/q$. 

Under these approximations, the complete set of EoMs reads
\begin{equation} \label{eq_final_m}
\begin{split}
\dot{m}^{\mu}=&  -\frac{\gamma}{2}m^{\mu}\left\lbrace 1+2 \braket {\braket{\mathrm{Re}[\xi^{\mu}f_{2}(\lbrace \xi \rbrace)] }}\right\rbrace \\
& -\gamma \braket {\braket{\mathrm{Re}[\xi^{*,\mu}f_{3}(\lbrace \xi \rbrace)] }} \\ 
&- 3\lambda \left\lbrace[1 -\mathrm{Re}(\omega)]y^{\mu}+\mathrm{Im}(\omega)\bar{y}^{\mu} \right\rbrace,\\
 \dot{x}^{\mu}=& -\frac{\gamma}{3} x^{\mu} - \lambda [(\mathrm{Re}(\omega^2)-1)y^{\mu} + \mathrm{Im}(\omega^2)\bar{y}^{\mu}], \\
 \dot{\bar{x}}^{\mu}=& -\frac{\gamma}{3} \bar{x}^{\mu} - \lambda [(\mathrm{Re}(\omega^2)-1)\bar{y}^{\mu} - \mathrm{Im}(\omega^2)y^{\mu}] \\
 \dot{y}^{\mu}= & -\frac{\gamma}{3} y^{\mu} + \lambda [(\mathrm{Re}(\omega^2)-1)x^{\mu} - \mathrm{Im}(\omega^2)\bar{x}^{\mu} +m^{\mu}], \\
 \dot{\bar{y}}^{\mu}= & -\frac{\gamma}{3} \bar{y}^{\mu} + \lambda [(\mathrm{Re}(\omega^2)-1)\bar{x}^{\mu} + \mathrm{Im}(\omega^2)x^{\mu} +\frac{m^{\mu}}{\sqrt{3}}].
\end{split}
\end{equation}
Note that, by getting the EoMs for the collective operators \eqref{eq_x_y_collective} by exploiting Eqs.~\eqref{eq_eomX_Y}, we first replace the damping term $ h_i^s(\alpha) $ by its average over the pattern disorder distribution, $h_i^s(\alpha) \rightarrow \braket{\braket{h^s(\alpha)}} = \frac{2}{3}$ (see Appendix \ref{Appendix B} for more details). 

Before going ahead with the analysis of the quantum qPHN as described by Eqs. \eqref{eq_final_m}, it is worth commenting upon their structure. First of all, all the $5p$ differential equations \eqref{eq_final_m} display a damping term, which will enable the system to reach a stationary state that we will characterize in the next section. Moreover, we can see that the EoMs of the overlaps and the ones of the coherences are coupled via terms which depend on the coherent control parameter $\lambda$. Such a feature already suggests that the addition of the quantum Hamiltonian \eqref{eq_Hterm} to the classical qPHN may play a non-trivial role with respect to the retrieval properties of the network, as we are going to further highlight in the remaining part of this work.
\section{Results}

In this section we investigate the long-time behavior of a quantum qPHN which evolves under Eqs.~\eqref{eq_final_m}, hereafter setting $\gamma=1$. We first focus on the case of one stored pattern, $p=1$. As pointed out in Sec.~\ref{Section 2}, a gauge transformation allows one to describe such single memory qPHN by means of the standard Potts model. For this reason, we eventually focus on the case $p=2$ (two memories), which displays a more interesting phenomenology.

\subsection{Purely dissipative case}
\label{Section VA}

\begin{figure}
\center
\includegraphics[width=\linewidth]{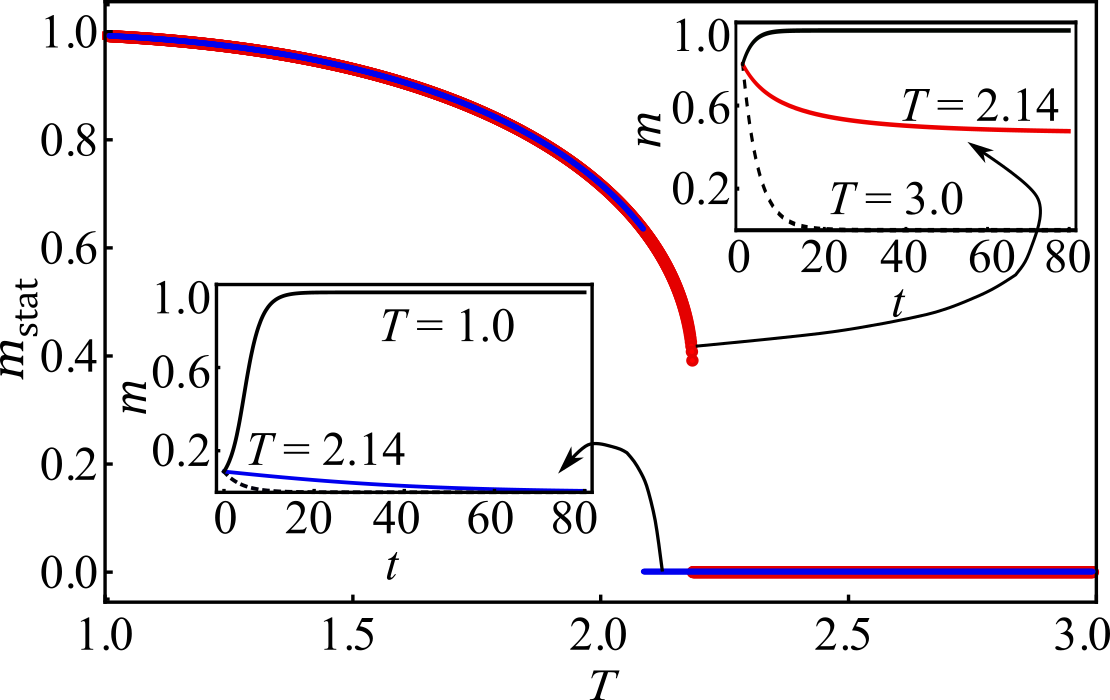}
\caption{\textbf{Phase diagram of the qPHN in the absence of quantum effects.} We set $q=3$, $\lambda=0$, $p=1$, $\gamma=1$. Stationary value of the overlap $m_{\mathrm{stat}}$ as a function of the temperature $T$. The values of the overlap are derived as long-time solutions of the EoMs, as explained in the text. Initial conditions have been chosen $ 0< m(0)<0.2$ for the red, thicker line, and $m(0)>0.5$ for the blue, thinner line. In both cases a paramagnetic phase, with a vanishing overlap, occurs at high temperature, and a retrieval phase, characterized by $m_{\mathrm{stat}} \approx 1$, takes place at low temperatures. The transition between the two phases occurs discontinuously, consistently with a first order phase transition. Furthermore, it is characterized by hysteresis: the transition from retrieval to paramagnetic occurs at lower temperatures if the initial condition is near the paramagnetic solution (blue, thinner line), and at higher temperatures if the initial condition is near the retrieval solution (red, thicker line). This feature is additionally stressed in the insets, which show the evolution of the overlap for different initial conditions. We set $m(0)>0.5$ for the top-right inset, and $0<m(0)<0.2$ for the bottom-left one. In both the insets, at the temperature $T=3$ (black dashed line) the overlap approaches the paramagnetic solution, and at $T=1$ (black solid line) the overlap approaches the retrieval solution. At the temperature $T=2.14$, the stationary solution depends on the initial condition, due to the coexistence of the retrieval and paramagnetic solutions, namely they are both stable fixed points of the dynamics. }\label{fig_pdiss_ov}
\end{figure}

Let us first consider the case $\lambda=0$, i.e. the absence of coherent effects, in order to establish a baseline and to investigate the role of temperature.
The results, displayed in Fig.~\ref{fig_pdiss_ov}, show the value of the overlap at sufficiently long times. By the latter expression we mean that our results are obtained from numerically integrating the EoMs \eqref{eq_final_m}, up to times such that the overlap converges to the same value, named $m_\mathrm{stat}$. We refer to this type of solution as long-time one, and it reasonably captures the behaviour of the system sufficiently close to stationarity. In Fig.~\ref{fig_pdiss_ov} the stationary solution is reached starting form an initial configuration that can partially overlap with the pattern, i.e. we select $m(0)\gtrsim 0$. Here we see that at low temperatures $m_{\mathrm{stat}} \approx 1$, signaling the retrieval of the pattern, whereas at high temperatures the overlap vanishes, $m_{\mathrm{stat}} \approx 0$, characterizing a paramagnetic phase. The transition between the two phases occurs discontinuously, consistently with a first order phase transition, as already highlighted in Fig.~\ref{fig_intro}(c). Furthermore, due to the existence of a coexistence region --- typical for first order transitions --- in the interval $T \in [T_m, T_M]$, the crossing is characterized by hysteresis, which manifests when starting from different initial conditions. Indeed, the red, thicker line in Fig.~\ref{fig_pdiss_ov} has been obtained as long-time solutions of the EoMs \eqref{eq_final_m} by choosing the initial condition near the stored pattern, $m(0) \approx 1$. In contrast, for the blue, thinner line we set $m(0) \approx 0$. 

The dependence on the choice of the initial conditions within the coexistence region is further highlighted in the insets. At relatively low and high temperatures, with respect to the critical one that emerges in the classical qPHN, ($T_c$ in Sec.~\ref{Section 2}), the results are consistent in both cases. To show an instance, we set $T=1$ and $T=3$ in the insets, highlighting retrieval of the pattern and non-retrieval, respectively. When setting the temperature to a value inside the coexistence region, e.g. $T=2.14$, the system shows retrieval of the pattern if the initial condition is close to the solution corresponding to the stored pattern (top right inset). Conversely, it displays non-retrieval if the initial condition is chosen so that the overlap between the network configuration and the pattern is almost vanishing (bottom left inset). 

\subsection{Quantum effects}

\begin{figure*}
\center
\includegraphics[width=\linewidth]{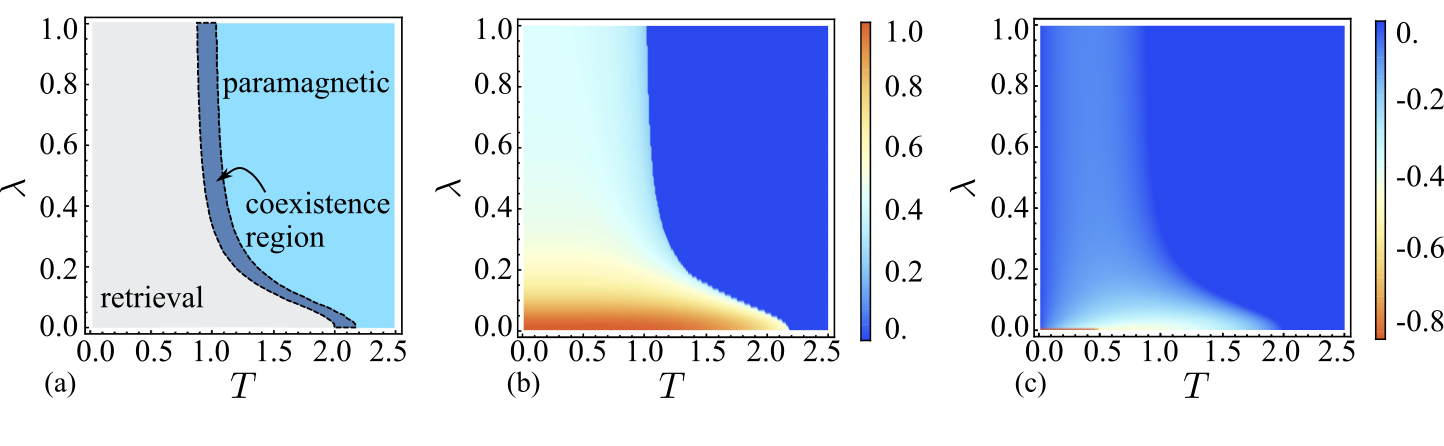}
\caption{\textbf{Phase diagram of the qPHN in the presence of thermal and quantum fluctuations.} We set $\gamma = 1$, $q=3$, and $p=1$. (a) Retrieval and paramagnetic phase. At high temperature (light blue region) there is one stationary solution of Eqs.~\eqref{eq_final_m} with vanishing order parameters, corresponding to a paramagnetic phase. At low temperature (grey region) the stationarity solutions are characterized by finite overlap values, both positive, $m > 0$, and negative, $m < 0$. This phase is referred to as retrieval phase. The dark blue region within dashed black lines divides paramagnetic phase and retrieval one, and it corresponds to a coexistence region. Here, both paramagnetic solution, $m=0$, and positive retrieval one, $m>0$, are present. (b, c) Stationary value of the overlap when the dynamics is initialized in the BA of the positive solution, and of the negative one, respectively. In the former case, the transition from the paramagnetic phase to the retrieval one is discontinuous, whereas in the latter case, such a transition is a continuous one.} \label{fig_llPD}
\end{figure*}

In the following we consider the long-time behaviour in the presence of quantum effects, i.e. $\lambda\neq 0$. The EoMs \eqref{eq_final_m} are numerically integrated at sufficiently long times, as explained at the beginning of Sec.~\ref{Section VA}. They thus provide us what we refer to as stationary solutions. In this case quantum fluctuations compete with thermal fluctuations, which gives rise to a rich phase diagram as shown in Fig.~\ref{fig_llPD} for the case $p=1$. The coherent control parameter $\lambda$ is here varied in the region $[0,1]$, which we refer to as "low" $\lambda$ regime. We will eventually consider the "high" $\lambda$ case, ($\lambda >1$), that gives rise to some additional physical effects. Referring to Fig.~\ref{fig_llPD}(a), at high temperature (light-blue region), we find one stationary solution of the EoMs. This is a paramagnetic solution, characterized by $m_{p}=y_{p}=\bar{y}_{p}=x_{p}=\bar{x}_{p}=0$. At low temperature (gray region), stationary solutions with finite overlap, $m \neq 0$, appear. We refer to this regime as retrieval phase. The two distinct phases are separated by a coexistence region (dark blue strip) of paramagnetic solution and positive retrieval solutions, $m>0$.

Also in the presence of quantum effects, the retrieval phase admits both positive and negative solutions, $m_{\mathrm{stat}}>0$, and $m_{\mathrm{stat}}<0$, respectively. This is highlighted in Fig.~\ref{fig_llPD}(b,c), where the transition from the paramagnetic to the retrieval phase is shown for varying $T$ and $\lambda$.  As commented in Sec.~\ref{Section 2}, negative solutions correspond to network configurations that are uncorrelated with respect to the pattern, at variance with the positive ones. These distinct fixed points are characterized by different basins of attraction (BAs). The latter determines which stationary solution will be reached, once the system is initialized within its BA. Thus, the value taken by the overlap towards stationarity, either signaling pattern retrieval ($m_{\mathrm{stat}}>0$) or not ($m_{\mathrm{stat}}<0$), depends on the chosen initial conditions. It is worth noticing that the transition between the paramagnetic solution and the retrieval one occurs discontinuously [see panel (b)], whereas it occurs continuously when negative solutions are reached [see panel (c)]. Such a phenomenology holds also in the purely dissipative case $\lambda=0$, where negative solutions vanish at $T=2$. Thus, consistently with our construction, the quantum qPHN reproduces the the classical phenomenology in absence of the Hamiltonian term~\eqref{eq_Hterm}. Finally, referring to the retrieval phase in panel (b), as the control parameter $\lambda$ is increased, the actual capacity of the network to store the pattern is diminished, as the overlap reduces from $m \approx 0.8 - 1.$ to $0.4$. 

\begin{figure}
\center
\includegraphics[width=\linewidth]{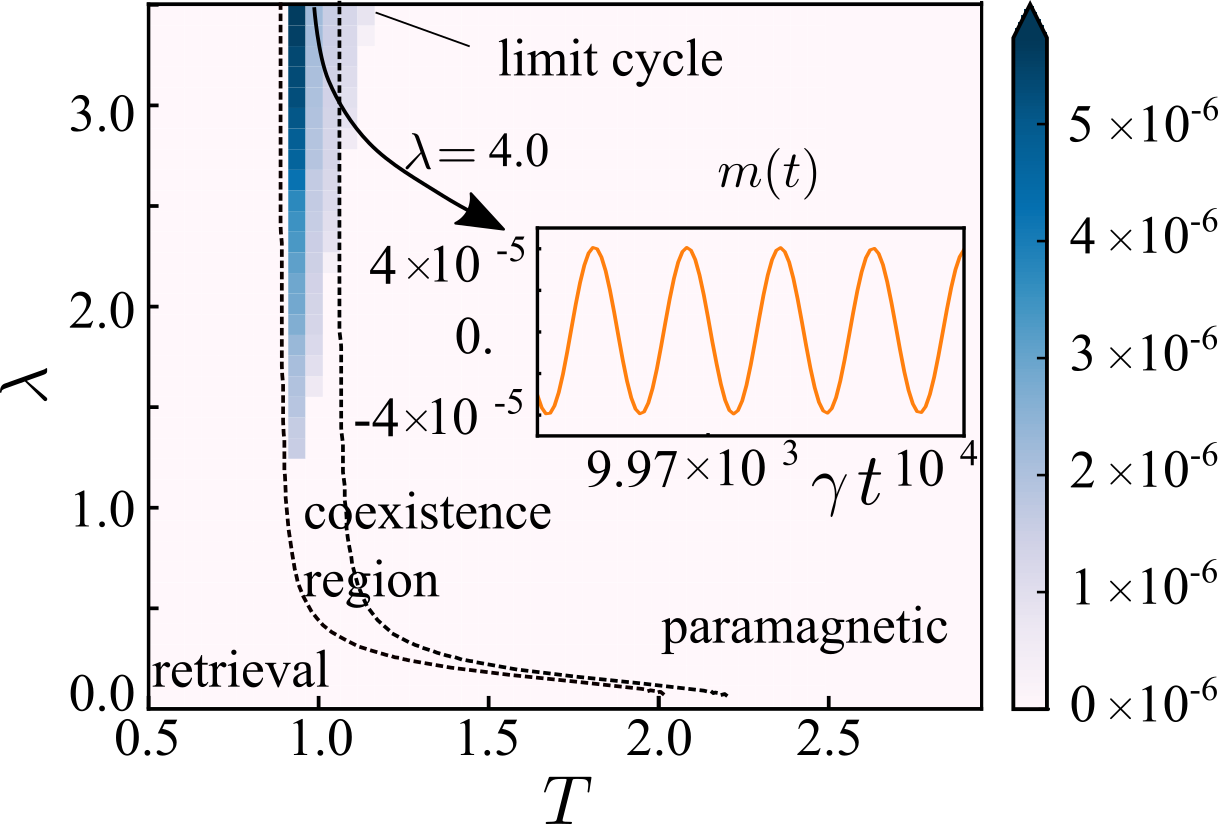}
\caption{\textbf{One memory limit cycle.} We set $\gamma = 1$, $q = 3$, $p = 1$. Standard deviation $\sigma_m$ of the overlap $m(t)$ at long times, with respect to the fixed point solutions, as defined by Eq.~\eqref{eq_standard_d}. Initial conditions are chosen such that $m(0)\approx 0 $. The dashed lines identify the separation among paramagnetic phase, coexistence region and retrieval phase. In the blue region $\sigma_m$ is finite, and such a portion of the phase diagram corresponds to a limit cycle phase. The inset shows the evolution of the overlap $m(t)$ at long times, highlighting an oscillatory behavior. We set $\lambda = 4.0$ , $T=0.86$.}\label{fig_LC1}
\end{figure}

\begin{figure}
\center
\includegraphics[width=\linewidth]{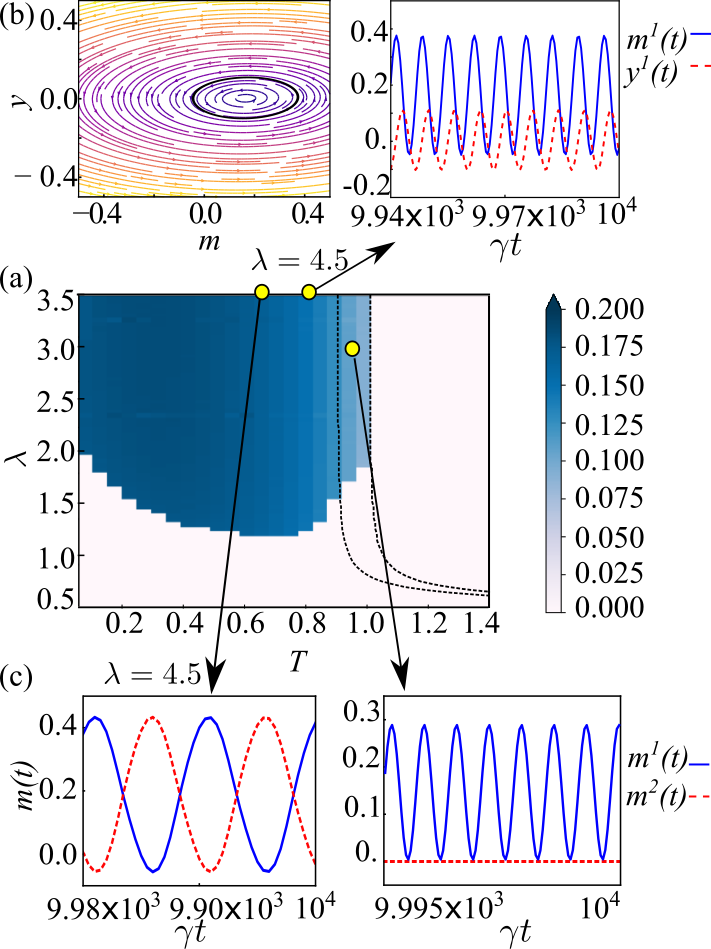}
\caption{\textbf{Two memory limit cycle phase.} We set $\gamma = 1$, $q = 3$, $p = 2$. (a) Standard deviation $\sigma_m$ of the overlap $m^{\mu}(t)$ at long times, with respect to the stationary solution. The dashed lines identify the separation among parmagnetic phase, coexistence region and retrieval phase. In the blue region $\sigma_m$ is finite, corresponding to a limit cycle phase. (b) The right-hand panel shows the time evolution of the overlap $m^{1}(t)$ and the coherence variable $y^{1}(t)$ at long times. In the left-hand panel the corresponding parametric plot and flux diagram are displayed, highlighting the closed orbit that characterizes the limit cycle solution. Colors for the flux diagram identify the norm of the corresponding vector field, which increases from purple to yellow. We set $T=0.8$ and $\lambda =4.5$. (c) Time evolution of the overlaps $m^{1}(t)$ and $m^{2}(t)$ as specified in the legend. We set $\lambda = 3.0$ and $T=0.99$ for the right-hand panel, where the self-sustained oscillations at long times appear for only one of the two overlap. For the left-hand panel it is $\lambda=4.5$ and $T=0.6$, and both the overlaps display a limit cycle solution. }\label{fig_LC2mem}
\end{figure}

From the phase diagram shown in Fig.~\ref{fig_llPD}, we see that the addition of the quantum coherent term to the classical dynamics still enables the network to retrieve patterns, but it diminishes its capability to fully recover patterns. Beyond that, quantum effects can, however, also lead to a qualitative change of the network's dynamics: for a certain range of parameters the overlap and the coherence variables display self-sustained oscillations at long times, as shown in Figs.~\ref{fig_LC1} and \ref{fig_LC2mem}. 
Let us now analyze in more detail these periodic long-time solutions, which occur in the "high"-$\lambda$ parameter regime and are referred to as limit cycle (LC) solutions. First we analyze the latter in the case $p=1$, and eventually we set $p=2$. The former case is highlighted in Fig.~\ref{fig_LC1}.  It shows the standard deviation $\sigma_m$ of the overlap $m(t_i)$ with respect to the time average of $m(t)$, say $\bar{m}$, at sufficiently long times. The values $m(t_i)$ are obtained at times $t_i \in I$, with $I$ an arbitrary time interval chosen at long times. Thus the quantity displayed reads 
\begin{equation}\label{eq_standard_d}
\sigma_m= \sqrt{\frac{1}{N_{I}}\sum_{i \in I} [m(t_i)-\bar{m}]^{2}}, 
\end{equation}
being $N_{I}$ the number of points $t_i \in I$. We set $N_{I}=2000$, and $I=[9 \times 10^3, 10^4]$. The light-blue region where $\sigma_m$ is finite corresponds to the LC phase. In the inset, we display the corresponding oscillating behavior of the overlap. As a matter of fact, the amplitude of the oscillations varies on a considerably small scale, $\sigma_m \lesssim 10^{-4} $.

As further detailed in App.~\ref{Appendix B}, we also explore the LC phase via a linear stability analysis. To this end, we linearize Eqs.~\eqref{eq_final_m} with respect to their stationary solutions, and study the eigenvalues of the corresponding Jacobian matrix. In this analysis we fix $\lambda$ to values well inside the region displaying a LC phase, and decrease the temperature $T$ starting from $T$-values corresponding to the paramagnetic phase. With changing $T$ and while approaching the LC phase, a complex conjugate pair of eigenvalues crosses the complex plane imaginary axes. In non-linear systems such a phenomenology is referred to as Hopf bifurcation \cite{Strogatz94}. It identifies a switching in the stability of the system, often giving rise to periodic solutions. For instance, Hopf bifurcations and LCs occur in the known Lotka-Volterra dynamical system \cite{Takeuchi96}, which is frequently used to model the dynamics of predator-prey interactions in biological systems \cite{NarendraEtAl71}. In two-dimensional dynamical systems, periodic long-time solutions characterizing LCs are displayed in phase space as isolated orbits \cite{Strogatz94}. In higher dimensional dynamical systems, as it is our case, one can still highlight isolated orbits by considering two out of the total number of dynamical variables involved, as we are going to show for the case of two stored patterns $p=2$.

Fig.~\ref{fig_LC2mem} (b), left-hand side panel, illustrates an instance of the isolated orbit that characterizes the LCs in our two-memories model. Here we show the parametric plot of the overlap $m^{1}(t)$ and its corresponding coherence variable $y^{1}(t)$. Trajectories surrounding the isolated orbit constitute the flux diagram of the vector field $(\dot{y}^{1},\dot{m}^{1})$, with the value of its norm color-coded and increasing from purple to yellow. We also set the parameter value inside the LC phase, e.g. $T=0.8$ and $\lambda=4.5$. Different choices of parameters lead to a similar phenomenology. 

The parameter region where the LC phase takes place is displayed in Fig.~\ref{fig_LC2mem}(a). It shows the standard deviation $\sigma_m$ defined by Eq.~\eqref{eq_standard_d}, averaged with respect to the two overlaps. At this point, it is worth considering the comparison with the LC phase of the single memory case shown in Fig.~\ref{fig_LC1}. We can see that the LC phase for the case $p=2$ is $(i)$ extended into the low temperature region, and $(ii)$ characterized by a larger amplitude of the oscillations occurring at long times. The latter feature can be further observed, e.g., from comparing the right-hand side of Fig.~\ref{fig_LC2mem}(b), which displays the time evolution of $m^{1}(t)$ and $y^{1}(t)$, and the inset of Fig.~\ref{fig_LC1}. Let us then try to provide a qualitative explanation of the features $(i)$ and $(ii)$. To this end, we contrast our results with those obtained for the quantum HNN analyzed in Ref.~\cite{Rotondo:JPA:2018}. There, similarly to our model, as a result of the competition between dissipation and coherent dynamics, a LC phase emerges. However, it turns out to be independent of the number of stored patterns. In other words, in this case there is no distinction between the case $p=1$ and $p=2$, at variance with our quantum qPHN model. A potential relevant difference between the HNN and the qPHN is the symmetry of the model. As already mentioned in Sec.~\ref{Section 2}, the former is characterized by a $\mathbb{Z}_2$ symmetry. Direct manifestation of the latter is that once the pattern $\lbrace \xi_1^{\mu},...,\xi_{N}^{\mu} \rbrace$ is stored as a memory, so is the opposite one, $\lbrace  - \xi_1^{\mu},...,- \xi_{N}^{\mu} \rbrace $. Contrarily, the $q \geq 2$ qPHN does not display the same symmetry, and the notion of "opposite" pattern is meaningless. Such a property might be the origin of the dissimilarity between the two models: when fixing $p=1$ and selecting the proper parameter regime, in the HNN the pattern and its opposite one can give rise to the LC phase. Instead, the $p=1$ qPHN cannot give rise to the same type of LC solutions, and it is only when considering at least $p=2$, that our model features two overlaps variables that can give rise to a LC phase. 

The behavior of the overlaps for a $p=2$ qPHN, and within the LC phase, is shown in Fig.~\ref{fig_LC2mem}(c). As visible in the right-hand side panel, close to the paramagnetic phase, there exists a narrow region where only one of the two patterns features a LC solution. Notably, the amplitude of the corresponding oscillations decreases with respect to the ones displayed in the left-hand side panel figure. Such a behavior takes place within the coexistence region of paramagnetic and retrieval fixed point solutions (the region within the dashed black lines). It is worth stressing that this region coincides also with the LC phase of the $p=1$ case (see Fig.~\ref{fig_LC1}). Moreover, also in this case the periodic solution is characterized by a relatively small amplitude of oscillations. These features hint that the coexistence of fixed point solutions may play a role in determining the properties of LC solutions occurring in such a parameter regime. In the bulk of the LC phase, as shown in the left-hand side panel of Fig.~\ref{fig_LC2mem}(c), the two overlaps present out-of-phase, self-sustained oscillations in the long-time limit. Here the amount of retrieval is relatively small, i.e.~it is at most $m \approx 0.4$. However, as a consequence of the out-of-phase oscillations between the two overlaps, the storage of either one of the two patterns is in principle possible. Indeed, the periodic switching between positive values of the two overlaps, with $m \lesssim 0.4$, implies that it is in principle possible to alternatively retrieve either one the two patterns. This phenomenology allows us to identify the bulk of the LC phase as a new type of "quantum" retrieval phase, emerging as a feature of the quantum formulation of the model.
%
\section{Conclusions and Outlook}

In this work, we have established a method for generalizing associative memory NNs via open quantum systems dynamics, which goes beyond the previously studied case of the HNN \cite{Rotondo:JPA:2018}. Starting point of our approach is the classical qPHN, whose non-equilibrium dynamics has been formulated in terms of a Markovian master equation of Lindblad form. This formulation allows to incorporate quantum fluctuations through a Hamiltonian term whose strength is given by a control parameter. By exploiting MF techniques, we construct the phase diagram for a many-body systems made of $N$ q-Potts spins. This shows that memory retrieval can be achieved for certain temperatures and values of this control parameter. Furthermore, the competition between quantum coherent dynamics and thermal fluctuations has shown to give rise to a new non-equilibrium quantum phase. The latter features LC solutions, which display persistent oscillations at long times. We analyze the LC phase for a single memory model and a two memories one, finding a different phenomenology in the two cases. The parameter regime where the LC phase occurs shrinks in the single memory case with respect to the two memories one. In the latter, both the overlaps sustain persistent, out-of-phase oscillations at long times which display a larger amplitude with respect to the one of the single memory case.

The analysis and characterization of the phase diagram done in this work represent initial steps in order to eventually explore more sophisticated questions. These concern, for example, the issue of identifying the storage capacity $\alpha_c$, i.e. how many patterns can be stored in a network with $N$ constituents. This aspect has been widely investigated in the classical realm. For instance, it is known that it is $\alpha_c= 0.14$ for a classical HNN \cite{AmitGS:1987}. For the case of a classical qPHN, the storage capacity depends on the number $q$ of levels of Potts spins, being $\alpha_c = 0.41$ if $q=3$ \cite{Kanter88, BolleDH_PRA_92, BolleDH_PA_92}. The main question here is understanding whether quantum formulations allow for a larger storage capacity than the classical models. One potential way to tackle such a problem has been pointed out in the recent theoretical work of Ref. \cite{LewensteinEtAl20}. Here, the authors consider a generic set of attractive quantum maps,
evolving input quantum states towards stationary ones, labeled as retrieval states. The size of such a stationary manifold gives access to the storage capacity, which is found to exponentially outperform the classical counterpart.
Our work lays the ground to tackle the question of storage capacity for the concrete and novel class of qPHNs introduced in our work, and to also investigate how physically realizable instances of such quantum neural networks can be engineered in state-of-the-art quantum hardware.

The presence of the novel types of retrieval phases identified in our work, which stem from the quantum description of the NNs, opens interesting new directions in relation to more quantum-information oriented questions: A first interesting question is how to formulate non-classical patterns, i.e.~specific input quantum states, in order to store and access them. Here, the investigation of the speed of retrieval \cite{Fiorelli:PRA:2019} with respect to the classical counterpart represents an important aspect. Furthermore, it would be worth exploring the link of quantum formulations of associative memories as realized by our quantum neural networks, with the active research field of quantum error correction. Indeed, one could investigate the capabilities of quantum generalization of Hopfield-type NNs in protecting and correcting faulty quantum states.

\section*{Acknowledgments}
We gratefully acknowledge useful discussions with L. B\"odeker. EF and MM acknowledge support by the ERC Starting Grant QNets Grant Number 804247. IL acknowledges support from the ``Wissenschaftler R\"{u}ckkehrprogramm GSO/CZS" of the Carl-Zeiss-Stiftung and the German Scholars Organization e.V. EF, MM and IL are grateful for funding from the Deutsche Forschungsgemeinschaft through Grant No. 449905436.
\appendix

\section{Equation of motion for the purely dissipative case}

In this section we derive the EoMs for the purely dissipative case ($\lambda = 0$) of the qPHN, showing that one can recover the HNN model for $q=2$.

As a first step, we evaluate the EoMs of the Potts spin operator $\hat{\Omega}_i^{\alpha}$, $i =0,...,N$. To this end, we exploit Eqs.~\eqref{eq_Zqalgebra1}-\eqref{eq_Zqalgebra3} for getting the following relations for the operators $\hat{T}^{i}_{\eta, \pm}$,
\begin{eqnarray}\label{eq2_subZq}
& \hat{T}^{i}_{\eta, \pm}\hat{\Omega}_{i}^{\alpha}=\omega^{\pm \alpha} \hat{\Omega}_{i}^{\alpha} \hat{T}^{i}_{\eta, \pm},\\ \label{eq2_proj}
& \hat{T}^{i}_{\eta, \pm}\hat{T}^{i}_{\eta, \mp}= \hat{P}^{i}_{\eta},\\ \label{eq2_sumProj}
& \sum_{\eta=1}^{q} \hat{T}^{i}_{\eta, \pm}\hat{T}^{i}_{\eta, \mp} = \mathbf{1}_{q},
\end{eqnarray}
where $\hat{P}_{\eta}= \ket{\eta-1}\bra{\eta-1}$.
The equations of motion read
\begin{equation}\label{e2_firstEoM}
\dot{\hat{\Omega}}_{i}^{\alpha}= \hat{\Omega}_i^{\alpha} \left\lbrace -  \gamma [(1- \omega^{-\alpha}) \hat{W}_{i,+} +(1-\omega^{\alpha}) \hat{W}_{i,-} ]\right\rbrace,
\end{equation}
where we have defined
\begin{eqnarray}
\gamma \hat{W}_{i,s} &\equiv& \sum_{\alpha=1}^{q} \hat{L}_{\alpha, s}^{i, \dagger}\hat{L}_{\alpha,s}^{i} = \gamma \sum_{\alpha=1}^{q} \Gamma_{\alpha,s}^{i,2}(\hat{T}_{\alpha,s}^{i})^{\dagger}\hat{T}_{\alpha,s}^{i} \label{e2_sumL1} \\
& =& \gamma \sum_{\alpha=1}^{q} \Gamma_{\alpha,s}^{i,2} \hat{P}^{i}_{\alpha + s} = \gamma\sum_{\alpha=1}^{q} \Gamma_{\alpha + s, s}^{i,2}\hat{P}_{\alpha}^{i}.\nonumber
\end{eqnarray}
It is straightforward to express the Potts spin operators $\hat{\Omega}_i^{\alpha}$ in terms of the projection operators as $\hat{\Omega}_{i}^{\alpha}=\sum_{\eta=1}^{q}\omega^{\alpha(\eta -1 )}\hat{P}_{\eta}^{\alpha}$. This relation can in general be inverted, yielding
\begin{equation}
\hat{P}_{\eta}^{i}=\sum_{a=1}^{q}g_{\eta a}\hat{\Omega}_{i}^{a}.
\end{equation}
with some coefficients $g_{\eta \alpha}$ that can be derived for each choice of $q$. The Potts spin operators evolve according to the equation
\begin{equation}
\begin{split}
&  \dot{\hat{\Omega}}_{i}^{\alpha} = -\gamma\sum_{a=1}^{q}f_{\alpha, a}^{i} \hat{\Omega}_{i}^{a}, \\
& f_{\alpha, a}^{i}= \sum_{\eta=1}^{q}g_{\eta a}\omega^{(\eta-1)\alpha}[(1-\omega^{-\alpha})\Gamma_{\eta-1,+}^{i,2}+(1-\omega^{\alpha})\Gamma_{\eta+1,-}^{i,2})].
\end{split}
\end{equation}

\subsubsection{The case q=2 - Hopfield neural network}

Taking $q=2$, the operator $\hat{\Omega}_i^{\alpha}$ becomes the spin $1/2$ Pauli operator $\hat{\sigma}^z_i$. Specializing Eq.~\eqref{e2_firstEoM} to this case yields
\begin{equation}\label{e2_caseq2}
\dot{\hat{\sigma}}^z_{i}=-2\gamma \hat{\sigma}^z_{i}(\hat{W}_{i,+}+\hat{W}_{i,-}),
\end{equation}
which can be further simplified as
\begin{eqnarray}
\hat{W}_{i,+} + \hat{W}_{i,-} & = & \sum_{\alpha=1}^{2}(\Gamma_{\alpha-1,+}^{i,2}+\Gamma_{\alpha+1,-}^{i,2})\hat{P}_{\alpha} \\
& = &  \frac{1}{Z_i}\sum_{\alpha=1}^{2}(e^{\beta \Delta \hat{E}_{\alpha-1,+}^{i}} + e^{\beta \Delta \hat{E}_{\alpha+1,-}^{i}})\hat{P}_{\alpha} \nonumber\\
& =  & \frac{1}{Z_i} (\Gamma_{2,+}^{i,2}+\Gamma_{2,-}^{i,2})\hat{P}_{1}+(\Gamma_{1,+}^{i,2}+\Gamma_{1,-}^{i,2})\hat{P}_{2}, \nonumber
\end{eqnarray}
where $\hat{P}_{1,2}=(1 \pm \hat{\sigma}_i^{z})/2$, and 
$\Gamma_{2,+}^{i,2}=\Gamma_{2,-}^{i,2}=e^{\beta \Delta \hat{E}_{H}}/Z$, $\Gamma_{1,+}^{i,2}=\Gamma_{1,-}^{i,2}=e^{-\beta \Delta \hat{E}_{H}}/Z$, and $Z=2(e^{-\beta \Delta \hat{E}_{H}}+e^{\beta \Delta \hat{E}_{H}})$. Thus, we get the equation for the Hopfield NN \cite{Rotondo:JPA:2018},
\begin{equation}\label{eq2_hopfieldZ}
\dot{\hat{\sigma}}_i^z=-\gamma \hat{\sigma}_i^z+\gamma \tanh{(\beta \Delta\hat{E}_H)}
\end{equation}
with $\Delta\hat{E}_H=\frac{1}{N}\sum_{j\neq i}J_{ij}\hat{\sigma}_j^z$. 

\section{Equation of motion for the q=3 Quantum Potts Neural Network}
\label{Appendix B}

This appendix contains details on how to derive Eqs.~\eqref{eq_final_m} of the main text, the quantum generalization of the $q=3$ Potts-Hopfield neural network. 

We proceed by specifying Eq.~\eqref{eq_Lindblad_op} for the operators $\hat{\Omega}_i$ and $\hat{\Omega}_{i}^{2} = \hat{\Omega}_{i}^{\dagger}= \hat{\Omega}_{i}^{-1}$. To this end, it is useful to write the projectors $\hat{P}_{\eta}=\ket{\eta-1}\bra{\eta-1}$, $\eta=1,2,3$ in terms of $\hat{\Omega}_i$, $\hat{\Omega}_i^{\dagger}$, and the identity operator $\mathbf{1}_3$ as follows
\begin{equation}\label{Ab_projectors}
\begin{split}
& \hat{P}^i_1=\frac{1}{(\omega-1)^2}[-\omega \hat{\Omega}_i+(1+\omega^2)\hat{\Omega}^{\dagger}_i-\omega \mathbf{1}_3],\\
& \hat{P}^i_2=\frac{1}{(\omega-1)^2}[- \hat{\Omega}_i+(1+\omega)\hat{\Omega}^{\dagger}_i-\omega \mathbf{1}_3],\\
& \hat{P}^i_3=\frac{1}{(\omega-1)^2}[(1+\omega) \hat{\Omega}_i-\hat{\Omega}^{\dagger}_i-\omega \mathbf{1}_3].\\
\end{split}
\end{equation}

Let us first derive the EoMs for Potts spin operators $\hat{\Omega}_i$, $\hat{\Omega}_i^{\dagger}$  without the Hamiltonian term, i.e. taking $\lambda = 0$. The evolution is governed by
\begin{equation}
\begin{split}
& \dot{\hat{\Omega}}_{i}=-\gamma[(\frac{1}{2}+\hat{f}_{1}^i)\hat{\Omega}_i+\hat{f}_{2}^i\hat{\Omega}_i^{\dagger}+\hat{f}_{3}^i],\\
\end{split}
\end{equation}
where we have defined
\begin{equation}\label{eq_appB_f1}
 \hat{f}_{1}^{i} = i\frac{1}{2 \sqrt{3}}[\hat{\Gamma}_{1,+}^{i,2}+\hat{\Gamma}_{2,+}^{i,2}+\hat{\Gamma}_{3,+}^{i,2}-(\hat{\Gamma}_{1,-}^{i,2}+\hat{\Gamma}_{2,-}^{i,2}+\hat{\Gamma}_{3,-}^{i,2})],
\end{equation}
\begin{equation}\label{eq_appB_f2}
\begin{split}
 \hat{f}_{2}^i & =  \frac{1}{2}[\hat{\Gamma}_{3,+}^{i,2}-\hat{\Gamma}_{3,-}^{i,2}-(\hat{\Gamma}_{2,+}^{i,2}-\hat{\Gamma}_{2,-}^{i,2})] \\
& +\frac{i}{2\sqrt{3}}[\hat{\Gamma}_{3,+}^{i,2}-\hat{\Gamma}_{3,-}^{i,2}+\hat{\Gamma}_{2,+}^{i,2}-\hat{\Gamma}_{2,-}^{i,2}-2(\hat{\Gamma}_{1,+}^{i,2}-\hat{\Gamma}_{1,-}^{i,2})],
\end{split}
\end{equation}
\begin{equation}\label{eq_appB_f3}
\begin{split}
\hat{f}_{3}^i &  = \frac{1}{2}(\hat{\Gamma}_{3,+}^{i,2}+\hat{\Gamma}_{2,-}^{i,2}-\hat{\Gamma}_{1,+}^{i,2}-\hat{\Gamma}_{1,-}^{i,2})  \\
& + \frac{i}{2\sqrt{3}}[\hat{\Gamma}_{3,+}^{i,2}-\hat{\Gamma}_{2,-}^{i,2}+\hat{\Gamma}_{1,+}^{i,2}-\hat{\Gamma}_{1,-}^{i,2}-2(\hat{\Gamma}_{2,+}^{i,2}-\hat{\Gamma}_{3,-}^{i,2})].\\
\end{split}
\end{equation}
The rates $\hat{\Gamma}_{\alpha,s}^{i}$ are given by Eq.~\eqref{eq_newJumps} and depend on the energy difference
\begin{equation}
\begin{split}
& \Delta \hat{E}_{\alpha,s}^{i} 
\\
&  = \sum_{\mu=1}^{p}\hat{m}^{\mu} \left\lbrace 3\cos[\frac{2 \pi }{3}(k_{i}^{\mu}-\alpha+1)] \right. \\
&  \left. \pm \sqrt{3}\sin[\frac{2 \pi }{3}(k_{i}^{\mu}-\alpha+1)]\right\rbrace  \\
& = 3\sum_{\mu=1}^{p}\hat{m}^{\mu}(\delta_{k_{i}^{\mu},\alpha-1}-\delta_{k_{i}^{\mu}\pm 1,\alpha-1}),
\end{split}
\end{equation}
for $\alpha=1,2,3$.

Let us now include the Hamiltonian term \eqref{eq_Hterm}, setting $\lambda \neq 0$. This generates a new term in the EoMs of the Potts operators which reads 
\begin{equation}
\begin{split}
i[\hat{H},\hat{\Omega}_{i}]& = - \lambda\sum_{\alpha=1}^{3} \omega^{\alpha-1}(\omega-1) \hat{Y}_{\alpha}^{i}.
\end{split}
\end{equation}
Here we have used that $\hat{\Omega}_{i}\hat{T}_{\alpha,+}^{i}=\sum_{\alpha'=1}^{3}\omega^{\alpha' -1 }\hat{T}_{\alpha',+}^{i}$ and $\hat{\Omega}_{i}\hat{T}_{\alpha,-}^{i}=\sum_{\alpha'=1}^{3}\omega^{\alpha' -1 } \omega (\hat{T}_{\alpha',+}^{i})^{\dagger}$, and we have defined the operators
\begin{equation}\label{eq2_generalizedY}
\hat{Y}_{\alpha}^{i}= \frac{1}{i}[\hat{T}_{\alpha,+}^{i}-(\hat{T}_{\alpha,+}^{i})^{\dagger}].
\end{equation}
Thus the full EoM of the Potts operators $\hat{\Omega}_{i}$ reads
\begin{equation}\label{App_eq2_omega_coherent}
\dot{\hat{\Omega}}_{i} = -\frac{\gamma}{2}\hat{\Omega}_{i} -\gamma[\hat{f}_{1}^{i}\hat{\Omega}_{i}+\hat{f}_{2}^{i}\hat{\Omega}^{\dagger}+\hat{f}_{3}^{i}]-\lambda \sum_{\alpha=1}^{3}\omega^{\alpha-1}(\omega-1)\hat{Y}_{\alpha}^{i}.
\end{equation}
To construct a closed set of EoM, we also need to derive the ones of the operators $\hat{Y}_{\alpha}^{i}$, and the ones of the operators
\begin{eqnarray} \label{eq2_generalizedX}
& \hat{X}_{\alpha}^{i}= \hat{T}_{\alpha,-}^{i}+(\hat{T}_{\alpha,-}^{i})^{\dagger}.
\end{eqnarray}
We first derive the EoMs of the operator $\hat{T}_{\alpha, s}^{i}$, which are more involved in comparison to the EoMs of the Potts spin operator $\hat{\Omega}_{i}$. This is due to the fact that the operators $\hat{T}_{\alpha, s}^{i}$ do not commute with the rates $\hat{\Gamma}_{\alpha',s'}^{j}$ that appear in the jumps operators~\eqref{eq_newJumps}. This gives rise to a cross-term, with $j \neq i$, in the Lindblad equation~\eqref{eq_Lindblad_op}, which reads
\begin{equation}
\sum_{\alpha',s'} \sum_{j \neq i} L_{\alpha',s'}^{j, \dagger}\hat{T}_{\alpha, s}^{i} L_{\alpha',s'}^{j} - \frac{1}{2}\lbrace L_{\alpha',s'}^{j, \dagger}L_{\alpha',s'}^{j}, \hat{T}_{\alpha, s}^{i}  \rbrace.
\end{equation}
For simplifying such an expression, it is sufficient to consider the non-commuting terms $\hat{\Gamma}_{\alpha', s'}^{j}\hat{T}_{\alpha, s}^{i}$ and exploit the relation~\eqref{eq2_subZq} for moving all the operators $\hat{T}_{\alpha, s}^{i}$ to the left, obtaining
\begin{equation}\label{eq_crossing term}
\begin{split}
& \hat{\Gamma}_{\alpha', s'}^{j}(\hat{\Omega}_{1},...,\hat{\Omega}_{i},..., \hat{\Omega}_{N})\hat{T}_{\alpha, s}^{i} \\
& = \hat{T}_{\alpha, s}^{i} \hat{\Gamma}_{\alpha', s'}^{j}(\hat{\Omega}_{1},...,\omega^{-s}\hat{\Omega}_{i},..., \hat{\Omega}_{N}) = \hat{T}_{\alpha, s}^{i} \hat{\Gamma}_{\alpha', s'}^{j (i)},
\end{split}
\end{equation}
with $\hat{\Gamma}_{\alpha', s'}^{j}(\hat{\Omega}_{1},...,\hat{\Omega}_{i},..., \hat{\Omega}_{N})$ highlighting the dependence of $\Gamma_{\alpha',s'}^{j}$ on all the $N$ Potts spin operators, and the compact notation $ \hat{\Gamma}_{\alpha', s'}^{j}(\hat{\Omega}_{1},...,\omega^{-s}\hat{\Omega}_{i},..., \hat{\Omega}_{N}) = \hat{\Gamma}_{\alpha', s'}^{j (i)} $ has been introduced in the second line. Now we can recognize that $  \hat{\Gamma}_{\alpha', s'}^{j} $ and $ \hat{\Gamma}_{\alpha', s'}^{j (i)}$ depend on two configurations which differ by a single Potts spin (the $i$-th), from which we expect that $  \hat{\Gamma}_{\alpha', s'}^{j} \approx \hat{\Gamma}_{\alpha', s'}^{j (i)}$ up to finite-size corrections which scale as $1/N$. As derived in Ref.~\cite{Rotondo:JPA:2018}, the crossing term~\eqref{eq_crossing term} in the EoMs for operators $\hat{T}_{\alpha,s}^{i}$ should more precisely scale as $O(p/N)$. It can be thus neglected for a subextensive number of pattern, $p \ll N$, and in the thermodynamic limit, leading to the EoM
\begin{equation}
\dot{\hat{T}}_{\alpha,s}^{i}= -\frac{\gamma}{2}\hat{T}_{\alpha,s}^{i}h^{s}_{i}(\alpha)+i\lambda[\hat{T}^{i}_{\alpha+2s,-s}+\hat{T}^{i}_{\alpha,-s}+\hat{P}^{i}_{\alpha+s}-\hat{P}_{\alpha}^{i}].
\end{equation}
Here $h^{s}_{i}(\alpha)=\sum_{s'=\pm}(\Gamma_{\alpha-s',s'}^{i,2}+\Gamma_{\alpha-s'+s,s'}^{i,2})$, and the operator $\hat{P}_{\eta}^i$ are defined by Eq.~\eqref{Ab_projectors} in terms of the Potts operator.

It follows that
\begin{equation}\label{eq2_eqX}
\begin{split}
& \dot{\hat{X}}_{\alpha}^{i}=  -\frac{\gamma}{2}h_{i}^{-}(\alpha)\hat{X}_{\alpha}^{i}-\lambda( \hat{Y}_{\alpha-2}^{i}- \hat{Y}_{\alpha}^{i}), \\ 
&  \dot{\hat{Y}}_{\alpha}^{i}=-\frac{\gamma}{2}h_{i}^{+}(\alpha)\hat{Y}_{\alpha}^{i}+ \lambda [\hat{X}_{\alpha +2}^{i}-\hat{X}_{\alpha}^{i}+2(\hat{P}_{\alpha+1}-\hat{P}_{\alpha})].
\end{split}
\end{equation}

It is worth noticing that, although Eqs.~\eqref{App_eq2_omega_coherent} and~\eqref{eq2_eqX} form a closed set of EoMs, they describe microscopic quantities, i.e.~individual Potts spins. Our interest lies, however, in understanding the behavior of macroscopic, collective observables, such as the overlap operators. When analyzing such quantities the number of equations is reduced from $O(N)$ to $O(p)$. 

To achieve such a reduction, we begin by writing EoMs for the overlaps
\begin{equation}\label{eq2_overlap_with_c}
\begin{split}
 \dot{\hat{m}}^{\mu}  &= -\frac{\gamma}{2}\hat{m}^{\mu}-\frac{\gamma}{2N} \sum_{i=1}^{N} [(\xi_{i}^{*,\mu}f_{1}^{i}+\xi_{i}^{\mu}\hat{f}^{i, \dagger}_{2})\hat{\Omega}_{i}+ \mathrm{h.c.}] \\
&- \frac{\gamma}{N}\sum_{i=1}^{N}\mathrm{Re}[\xi_{i}^{*,\mu}\hat{f}_{3}^{i}] \\
& -\frac{\lambda}{2N}\sum_{i=1}^{N}\sum_{\alpha=1}^{3}\hat{Y}_{\alpha}^{i}[\xi_{i}^{*, \mu}\omega^{\alpha-1}(\omega-1)+\xi_{i}^{\mu}\omega^{-(\alpha-1)}(\omega^{*}-1)].
\end{split}
\end{equation}
The last term suggests to also consider the collective operators
\begin{equation}
\begin{split}
& 
\hat{x}^{\mu}=\frac{1}{6N}\sum_{\alpha=1}^{3}  \sum_{i=1}^{N} \omega^{\alpha}\xi_{i}^{\mu} \hat{X}_{\alpha}^{i} + \mathrm{h. c. }, \\
&  \hat{\bar{x}}^{\mu}=i \left[\frac{1}{6N}\sum_{\alpha=1}^{3}\sum_{i=1}^{N}\omega^{-\alpha}\xi_{i}^{*, \mu} \hat{X}_{\alpha}^{i} - \mathrm{h. c. } \right], \\
\end{split}
\end{equation}
and two additional ones, $\hat{y}^{\mu}$ and $\hat{\bar{y}}^{\mu}$, obtained by replacing $\hat{X}_{\alpha}^{i}$ with $\hat{Y}_{\alpha}^{i}$. This choice will allow us to find a closed set of EoMs, when providing the system with some additional approximations.

To achieve this, we firstly perform a mean field (MF) approximation, replacing collective operators with their averaged quantities, and neglecting correlations amongst them. For the classical system, MF is exact in the thermodynamic limit, and we expect such a treatment to be appropriate also in our quantum generalization \cite{Gayrard92, Rotondo:JPA:2018}. Thus, we derive EoMs for averaged quantities $O^{\mu}$ such that $\braket{O}^{\mu} \equiv O^{\mu} $. 

\begin{figure}
\includegraphics[width=\linewidth]{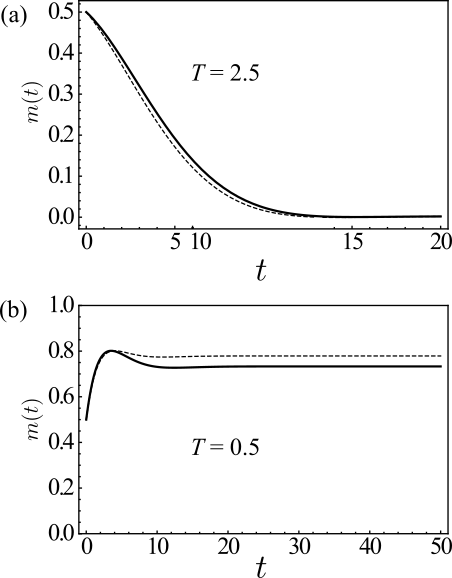}
\caption{\textbf{Approximated and exact dynamics.} We set $p=1$, $q=3$, $\gamma = 1$, $\lambda = 0.1$. Overlap $m(t)$ as a function of the time $t$ for (a) $T=2.5 $, and (b) $T=0.5$. Dashed black line corresponds to the exact solution, the solid black line represents the mean field, self-averaged case.}\label{fig_dumappr}
\end{figure}

Secondly, we expect the system near stationarity to be mostly determined by the behavior of the overlap. Indeed, we build up the model so that it exactly reproduces the classical case in the limit $\lambda \rightarrow 0$, and quantum effects are gradually taken into account as $\lambda$ is increased. Moreover, the classical case is well described by the dynamics of the overlap variables. We thus expect that also the description of the quantum dynamics near to the stationary solutions can be mainly
embodied in the dynamics of the overlaps. Therefore, we assume that approximately $\braket{\hat{\Omega}_{i}}\approx \xi_{i}^{\mu} m^{\mu}$ holds. This allows us to treat terms of the type $\sum_{i=1}^{N}\xi_{i}^{\mu}\hat{\Omega}_i \hat{f}_{i}^{ \dagger}$ which appear in the Eq.~\eqref{eq2_overlap_with_c} of the overlap. Furthermore, Eqs.~\eqref{eq2_eqX} give rise to a set of EoMs which do not immediately form a closed set for the collective operators $\hat{x}^{\mu}, \hat{y}^{\mu}$. This is due to the presence of terms of the type $ -\frac{1}{6N}\sum_{\alpha=1}^{3}\omega^{\alpha}\sum_{i=1}^{N} h_i^-(\alpha) \hat{X}_{\alpha}^{i}\xi_{i}^{\mu}$. Here, we exploit the average of the damping coefficient $h_i^s(\alpha)$ with respect to the pattern disorder distribution. An analysis of the dynamics obtained under this approximation and the dynamics without such an average is displayed in Fig.~\ref{fig_dumappr}, panel (a) and (b). Here we show the evolution of the overlap for a single pattern, $p=1$. The dashed black line is obtained by solving the exact dynamics for $N=2$, and the solid black line represents the approximated one. We fixed $\lambda = 0.1$ and $\gamma = 1$. Panel (a) and (b) show the result for $T=2.5$ and  $T=0.5$, respectively. At large temperatures, the difference between the two results occurs at early stages of the dynamics only, whereas at small temperatures, the performed approximations yields a constant, although relatively small shift that holds even at stationarity. However, the qualitative behavior of the network remains unaltered.

The final step towards obtaining the EoMs for the macroscopic observables consists in exploiting the so-called self-averaging property, typical of disordered systems \cite{Mezard:book}. This means, that in the large $N$ limit and for $p/N \rightarrow 0$, we use that $ \frac{1}{N}\sum_{i}g(\xi_{i}^{\mu}) \rightarrow \sum_{\lbrace \xi^{\mu} \rbrace} \mathbf{P}(\xi)g(\xi^{\mu}) \equiv \braket{\braket{g(\xi^{\mu})}}$. 

After the above approximations the EoMs read
\begin{equation} \label{AppB_final_m}
\begin{split}
\dot{m}^{\mu}=&  -\frac{\gamma}{2}m^{\mu}\left\lbrace 1+2 \braket {\braket{\mathrm{Re}[\xi^{\mu}f_{2}(\lbrace \xi \rbrace)] }}\right\rbrace \\
& -\gamma \braket {\braket{\mathrm{Re}[\xi^{*,\mu}f_{3}(\lbrace \xi \rbrace)] }} \\ 
&- 3\lambda \left\lbrace[1 -\mathrm{Re}(\omega)]y^{\mu}+\mathrm{Im}(\omega)\bar{y}^{\mu} \right\rbrace,\\
 \dot{x}^{\mu}=& -\frac{\gamma}{3} x^{\mu} - \lambda [(\mathrm{Re}(\omega^2)-1)y^{\mu} + \mathrm{Im}(\omega^2)\bar{y}^{\mu}], \\
 \dot{\bar{x}}^{\mu}=& -\frac{\gamma}{3} \bar{x}^{\mu} - \lambda [(\mathrm{Re}(\omega^2)-1)\bar{y}^{\mu} - \mathrm{Im}(\omega^2)y^{\mu}] \\
 \dot{y}^{\mu}= & -\frac{\gamma}{3} y^{\mu} + \lambda [(\mathrm{Re}(\omega^2)-1)x^{\mu} - \mathrm{Im}(\omega^2)\bar{x}^{\mu} +m^{\mu}], \\
 \dot{\bar{y}}^{\mu}= & -\frac{\gamma}{3} \bar{y}^{\mu} + \lambda [(\mathrm{Re}(\omega^2)-1)\bar{x}^{\mu} + \mathrm{Im}(\omega^2)x^{\mu} +m^{\mu}/\sqrt{3}],
\end{split}
\end{equation}
which are $5p$ differential equations for the collective operators. The presence of a damping term, proportional to $\gamma$,  enables the system to reach a stationary state. The characterization of the latter with varying of $\lambda$ and $T$ is performed in the main text. However, we can already see that the EoMs of the overlaps and the ones of the coherences are coupled via terms proportional to the coherent control parameter $\lambda$. This suggests that the addition of the quantum Hamiltonian~\eqref{eq_Hterm} to the classical qPHN may play a non-trivial role when analyzing the retrieval properties of the network.

\subsection{p=1 limit cycle phase}

\begin{figure}[H]
\center
\includegraphics[width=\linewidth]{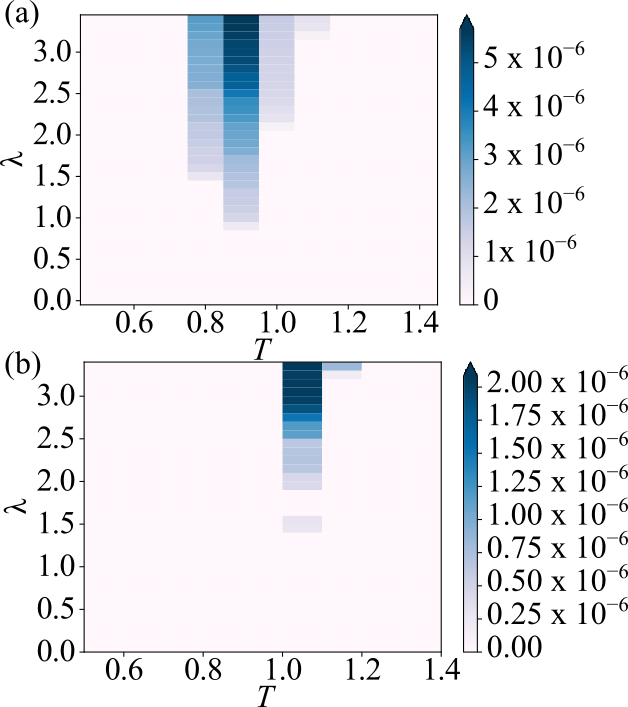}
\caption{\textbf{One memory LC phase.} We set $\gamma = 1$, $q = 3$, $p = 1$. Standard deviation $\sigma_m$ of the overlap $m(t)$ at long times, with respect to the fixed point solutions, as defined by Eq.~\eqref{eq_standard_d}. Initial conditions are chosen (a) $m(0)< 0 $ and (b) $m(0)>0$. In the blue region $\sigma_m$ is finite, and such a portion of the phase diagram corresponds to a limit cycle phase.}\label{fig_LC_A}
\end{figure}

At stationarity, Eqs.~\eqref{eq_final_m} display limit cycle solutions for a certain range of temperatures $T$ and when the control parameter $\lambda$ is sufficiently large. As shown in Fig.~\ref{fig_LC1} in the main text, we analyze the standard deviation of the overlap variables at long times, with respect to the fixed point solutions of the EoMs. Additional results are reported in Fig.~\ref{fig_LC_A}, where panels (a) and (b) show the standard deviation defined by Eq.~\eqref{eq_standard_d}, with initial conditions $m(0)<0$ and $m(0)>0$, respectively. The shaded region corresponds to a finite value of the standard deviation $\sigma_m$.

\begin{figure}[H]
\center
\includegraphics[width=\linewidth]{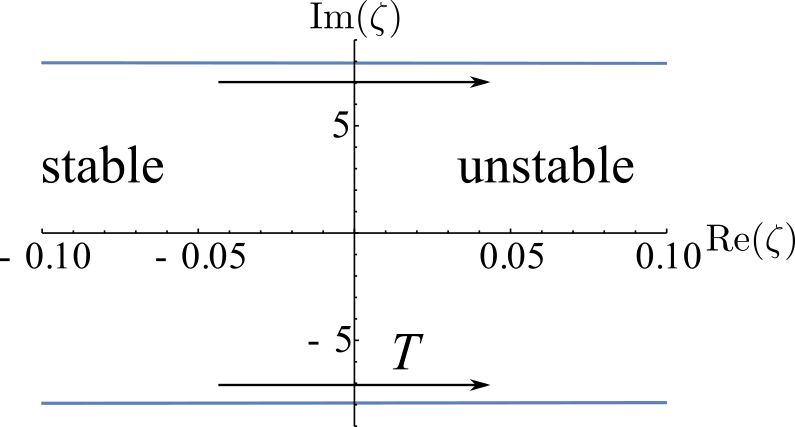}
\caption{ \textbf{Hopf bifurcation. } Complex plane representation of eigenvalues $\zeta_{1,2}$ of the linearized Eqs. \eqref{eq_final_m} with respect to the paramagnetic solution. We fix $\lambda = 2.5$ and the temperature is varied in the interval $T \in [0.5, 1.5]$. The change in sign of the eigenvalues while keeping a finite imaginary part signals a Hopf bifurcation. The paramagnetic solution is stable[unstable] for $\mathrm{Re}(\zeta_{1,2}) < [>] 0$. However, as the imaginary axis is crossed by lowering the temperature, a limit cycle solution appears, as discussed in the main text. Other parameters are $p=1$, $\gamma=1$. }\label{fig_LC2}
\end{figure}

Additionally, we perform a stability analysis of the fixed point paramagnetic solution. Namely, defining $\vec{v}=( m,  x ,  \bar{x},  y,  \bar{y})^{T}$, we linearize the Eqs.~\eqref{eq_final_m} with respect to the solution $\vec{v}=\vec{v}_{p} + \delta \vec{v}$, with $\vec{v}_{p}=0$. We obtain the equations in the form $\delta \dot{\vec{v}} = J \delta \vec{v}$, where $J$ is the Jacobian matrix. From the eigenvalues of the Jacobian matrix, say $\vec{\zeta}$, we could identify a Hopf bifurcation~\cite{Strogatz94}, signaling a limit cycle solution. An instance of this feature is shown in Fig.~\ref{fig_LC2}, which displays two complex conjugate eigenvalues of the Jacobian matrix, $\zeta_1$ and $\zeta_2$. The coherent control parameter is fixed to the value $\lambda = 2.5$, and the temperature is varied in the interval $T \in [0.5, 1.5 ]$. We can see that the real part of the eigenvalues changes sign when crossing the imaginary axis, i.e. with finite imaginary part, highlighting the feature of a Hopf bifurcation \cite{Strogatz94}.

\bibliography{DM_bib2}

\begin{thebibliography}{63}%
\makeatletter
\providecommand \@ifxundefined [1]{%
 \@ifx{#1\undefined}
}%
\providecommand \@ifnum [1]{%
 \ifnum #1\expandafter \@firstoftwo
 \else \expandafter \@secondoftwo
 \fi
}%
\providecommand \@ifx [1]{%
 \ifx #1\expandafter \@firstoftwo
 \else \expandafter \@secondoftwo
 \fi
}%
\providecommand \natexlab [1]{#1}%
\providecommand \enquote  [1]{``#1''}%
\providecommand \bibnamefont  [1]{#1}%
\providecommand \bibfnamefont [1]{#1}%
\providecommand \citenamefont [1]{#1}%
\providecommand \href@noop [0]{\@secondoftwo}%
\providecommand \href [0]{\begingroup \@sanitize@url \@href}%
\providecommand \@href[1]{\@@startlink{#1}\@@href}%
\providecommand \@@href[1]{\endgroup#1\@@endlink}%
\providecommand \@sanitize@url [0]{\catcode `\\12\catcode `\$12\catcode
  `\&12\catcode `\#12\catcode `\^12\catcode `\_12\catcode `\%12\relax}%
\providecommand \@@startlink[1]{}%
\providecommand \@@endlink[0]{}%
\providecommand \url  [0]{\begingroup\@sanitize@url \@url }%
\providecommand \@url [1]{\endgroup\@href {#1}{\urlprefix }}%
\providecommand \urlprefix  [0]{URL }%
\providecommand \Eprint [0]{\href }%
\providecommand \doibase [0]{http://dx.doi.org/}%
\providecommand \selectlanguage [0]{\@gobble}%
\providecommand \bibinfo  [0]{\@secondoftwo}%
\providecommand \bibfield  [0]{\@secondoftwo}%
\providecommand \translation [1]{[#1]}%
\providecommand \BibitemOpen [0]{}%
\providecommand \bibitemStop [0]{}%
\providecommand \bibitemNoStop [0]{.\EOS\space}%
\providecommand \EOS [0]{\spacefactor3000\relax}%
\providecommand \BibitemShut  [1]{\csname bibitem#1\endcsname}%
\let\auto@bib@innerbib\@empty
\bibitem [{\citenamefont {Samuel}(1959)}]{Samuel59}%
  \BibitemOpen
  \bibfield  {author} {\bibinfo {author} {\bibfnamefont {A.~L.}\ \bibnamefont
  {Samuel}},\ }\href {\doibase 10.1147/rd.33.0210} {\bibfield  {journal}
  {\bibinfo  {journal} {IBM Journal of Research and Development}\ }\textbf
  {\bibinfo {volume} {3}},\ \bibinfo {pages} {210} (\bibinfo {year}
  {1959})}\BibitemShut {NoStop}%
\bibitem [{\citenamefont {Goodfellow}\ \emph {et~al.}(2016)\citenamefont
  {Goodfellow}, \citenamefont {Bengio},\ and\ \citenamefont
  {Courville}}]{GoodfellowEtAl16}%
  \BibitemOpen
  \bibfield  {author} {\bibinfo {author} {\bibfnamefont {I.}~\bibnamefont
  {Goodfellow}}, \bibinfo {author} {\bibfnamefont {Y.}~\bibnamefont {Bengio}},
  \ and\ \bibinfo {author} {\bibfnamefont {A.}~\bibnamefont {Courville}},\
  }\href@noop {} {\emph {\bibinfo {title} {Deep Learning}}}\ (\bibinfo
  {publisher} {MIT Press},\ \bibinfo {year} {2016})\ \bibinfo {note}
  {\url{http://www.deeplearningbook.org}}\BibitemShut {NoStop}%
\bibitem [{\citenamefont {Jordan}\ and\ \citenamefont
  {Mitchell}(2015)}]{JordanM15}%
  \BibitemOpen
  \bibfield  {author} {\bibinfo {author} {\bibfnamefont {M.~I.}\ \bibnamefont
  {Jordan}}\ and\ \bibinfo {author} {\bibfnamefont {T.~M.}\ \bibnamefont
  {Mitchell}},\ }\href {\doibase 10.1126/science.aaa8415} {\bibfield  {journal}
  {\bibinfo  {journal} {Science}\ }\textbf {\bibinfo {volume} {349}},\ \bibinfo
  {pages} {255} (\bibinfo {year} {2015})}\BibitemShut {NoStop}%
\bibitem [{\citenamefont {LeCun}\ \emph {et~al.}(2015)\citenamefont {LeCun},
  \citenamefont {Bengio},\ and\ \citenamefont {Hinton}}]{LeCunEtAl15}%
  \BibitemOpen
  \bibfield  {author} {\bibinfo {author} {\bibfnamefont {Y.}~\bibnamefont
  {LeCun}}, \bibinfo {author} {\bibfnamefont {Y.}~\bibnamefont {Bengio}}, \
  and\ \bibinfo {author} {\bibfnamefont {G.}~\bibnamefont {Hinton}},\
  }\href@noop {} {\bibfield  {journal} {\bibinfo  {journal} {Nature}\ }\textbf
  {\bibinfo {volume} {521}},\ \bibinfo {pages} {436} (\bibinfo {year}
  {2015})}\BibitemShut {NoStop}%
\bibitem [{\citenamefont {Haykin}(1998)}]{Haykin98}%
  \BibitemOpen
  \bibfield  {author} {\bibinfo {author} {\bibfnamefont {S.}~\bibnamefont
  {Haykin}},\ }\href@noop {} {\emph {\bibinfo {title} {Neural Networks: A
  Comprehensive Foundation}}},\ \bibinfo {edition} {2nd}\ ed.\ (\bibinfo
  {publisher} {Prentice Hall PTR},\ \bibinfo {address} {USA},\ \bibinfo {year}
  {1998})\BibitemShut {NoStop}%
\bibitem [{\citenamefont {Crick}(1989)}]{Crick89}%
  \BibitemOpen
  \bibfield  {author} {\bibinfo {author} {\bibfnamefont {F.}~\bibnamefont
  {Crick}},\ }\href@noop {} {\bibfield  {journal} {\bibinfo  {journal}
  {Nature}\ }\textbf {\bibinfo {volume} {337}},\ \bibinfo {pages} {129}
  (\bibinfo {year} {1989})}\BibitemShut {NoStop}%
\bibitem [{\citenamefont {Amit}(1992)}]{Amit_book}%
  \BibitemOpen
  \bibfield  {author} {\bibinfo {author} {\bibfnamefont {D.~J.}\ \bibnamefont
  {Amit}},\ }\href@noop {} {\emph {\bibinfo {title} {Modelling Brain Function:
  The World of Attractor Neural Networks}}},\ \bibinfo {edition} {1st}\ ed.\
  (\bibinfo  {publisher} {Cambridge University Press},\ \bibinfo {address}
  {USA},\ \bibinfo {year} {1992})\BibitemShut {NoStop}%
\bibitem [{\citenamefont {Montanaro}(2016)}]{Montanaro16}%
  \BibitemOpen
  \bibfield  {author} {\bibinfo {author} {\bibfnamefont {A.}~\bibnamefont
  {Montanaro}},\ }\href {\doibase 10.1038/npjqi.2015.23} {\bibfield  {journal}
  {\bibinfo  {journal} {npj Quantum Information}\ }\textbf {\bibinfo {volume}
  {2}} (\bibinfo {year} {2016}),\ 10.1038/npjqi.2015.23}\BibitemShut {NoStop}%
\bibitem [{\citenamefont {Nielsen}\ and\ \citenamefont
  {Chuang}(2011)}]{NielsenC11}%
  \BibitemOpen
  \bibfield  {author} {\bibinfo {author} {\bibfnamefont {M.~A.}\ \bibnamefont
  {Nielsen}}\ and\ \bibinfo {author} {\bibfnamefont {I.~L.}\ \bibnamefont
  {Chuang}},\ }\href@noop {} {\emph {\bibinfo {title} {Quantum Computation and
  Quantum Information: 10th Anniversary Edition}}},\ \bibinfo {edition} {10th}\
  ed.\ (\bibinfo  {publisher} {Cambridge University Press},\ \bibinfo {address}
  {USA},\ \bibinfo {year} {2011})\BibitemShut {NoStop}%
\bibitem [{\citenamefont {Shor}(1999)}]{Shor:SIAM:1999}%
  \BibitemOpen
  \bibfield  {author} {\bibinfo {author} {\bibfnamefont {P.~W.}\ \bibnamefont
  {Shor}},\ }\href {\doibase 10.1137/S0036144598347011} {\bibfield  {journal}
  {\bibinfo  {journal} {SIAM Review}\ }\textbf {\bibinfo {volume} {41}},\
  \bibinfo {pages} {303} (\bibinfo {year} {1999})}\BibitemShut {NoStop}%
\bibitem [{\citenamefont {Grover}(1997)}]{Grover:PRL:1997}%
  \BibitemOpen
  \bibfield  {author} {\bibinfo {author} {\bibfnamefont {L.~K.}\ \bibnamefont
  {Grover}},\ }\href {\doibase 10.1103/PhysRevLett.79.325} {\bibfield
  {journal} {\bibinfo  {journal} {Phys. Rev. Lett.}\ }\textbf {\bibinfo
  {volume} {79}},\ \bibinfo {pages} {325} (\bibinfo {year} {1997})}\BibitemShut
  {NoStop}%
\bibitem [{\citenamefont {Carleo}\ \emph {et~al.}(2019)\citenamefont {Carleo},
  \citenamefont {Cirac}, \citenamefont {Cranmer}, \citenamefont {Daudet},
  \citenamefont {Schuld}, \citenamefont {Tishby}, \citenamefont
  {Vogt-Maranto},\ and\ \citenamefont {Zdeborov\'a}}]{CarleoEtAl19}%
  \BibitemOpen
  \bibfield  {author} {\bibinfo {author} {\bibfnamefont {G.}~\bibnamefont
  {Carleo}}, \bibinfo {author} {\bibfnamefont {I.}~\bibnamefont {Cirac}},
  \bibinfo {author} {\bibfnamefont {K.}~\bibnamefont {Cranmer}}, \bibinfo
  {author} {\bibfnamefont {L.}~\bibnamefont {Daudet}}, \bibinfo {author}
  {\bibfnamefont {M.}~\bibnamefont {Schuld}}, \bibinfo {author} {\bibfnamefont
  {N.}~\bibnamefont {Tishby}}, \bibinfo {author} {\bibfnamefont
  {L.}~\bibnamefont {Vogt-Maranto}}, \ and\ \bibinfo {author} {\bibfnamefont
  {L.}~\bibnamefont {Zdeborov\'a}},\ }\href {\doibase
  10.1103/RevModPhys.91.045002} {\bibfield  {journal} {\bibinfo  {journal}
  {Rev. Mod. Phys.}\ }\textbf {\bibinfo {volume} {91}},\ \bibinfo {pages}
  {045002} (\bibinfo {year} {2019})}\BibitemShut {NoStop}%
\bibitem [{\citenamefont {Carleo}\ and\ \citenamefont
  {Troyer}(2017)}]{CarleoT17}%
  \BibitemOpen
  \bibfield  {author} {\bibinfo {author} {\bibfnamefont {G.}~\bibnamefont
  {Carleo}}\ and\ \bibinfo {author} {\bibfnamefont {M.}~\bibnamefont
  {Troyer}},\ }\href {\doibase 10.1126/science.aag2302} {\bibfield  {journal}
  {\bibinfo  {journal} {Science}\ }\textbf {\bibinfo {volume} {355}},\ \bibinfo
  {pages} {602} (\bibinfo {year} {2017})}\BibitemShut {NoStop}%
\bibitem [{\citenamefont {Gao}\ and\ \citenamefont {Duan}(2017)}]{GaoD17}%
  \BibitemOpen
  \bibfield  {author} {\bibinfo {author} {\bibfnamefont {X.}~\bibnamefont
  {Gao}}\ and\ \bibinfo {author} {\bibfnamefont {L.-M.}\ \bibnamefont {Duan}},\
  }\href@noop {} {\bibfield  {journal} {\bibinfo  {journal} {Nature
  communications}\ }\textbf {\bibinfo {volume} {8}},\ \bibinfo {pages} {1}
  (\bibinfo {year} {2017})}\BibitemShut {NoStop}%
\bibitem [{\citenamefont {Huang}\ and\ \citenamefont {Moore}(2017)}]{HuangM17}%
  \BibitemOpen
  \bibfield  {author} {\bibinfo {author} {\bibfnamefont {Y.}~\bibnamefont
  {Huang}}\ and\ \bibinfo {author} {\bibfnamefont {J.~E.}\ \bibnamefont
  {Moore}},\ }\href@noop {} {\bibfield  {journal} {\bibinfo  {journal} {arXiv
  preprint arXiv:1701.06246}\ } (\bibinfo {year} {2017})}\BibitemShut {NoStop}%
\bibitem [{\citenamefont {Choo}\ \emph {et~al.}(2018)\citenamefont {Choo},
  \citenamefont {Carleo}, \citenamefont {Regnault},\ and\ \citenamefont
  {Neupert}}]{ChooEtAl18}%
  \BibitemOpen
  \bibfield  {author} {\bibinfo {author} {\bibfnamefont {K.}~\bibnamefont
  {Choo}}, \bibinfo {author} {\bibfnamefont {G.}~\bibnamefont {Carleo}},
  \bibinfo {author} {\bibfnamefont {N.}~\bibnamefont {Regnault}}, \ and\
  \bibinfo {author} {\bibfnamefont {T.}~\bibnamefont {Neupert}},\ }\href
  {\doibase 10.1103/PhysRevLett.121.167204} {\bibfield  {journal} {\bibinfo
  {journal} {Phys. Rev. Lett.}\ }\textbf {\bibinfo {volume} {121}},\ \bibinfo
  {pages} {167204} (\bibinfo {year} {2018})}\BibitemShut {NoStop}%
\bibitem [{\citenamefont {Saito}(2018)}]{Saito18}%
  \BibitemOpen
  \bibfield  {author} {\bibinfo {author} {\bibfnamefont {H.}~\bibnamefont
  {Saito}},\ }\href {\doibase 10.7566/JPSJ.87.074002} {\bibfield  {journal}
  {\bibinfo  {journal} {Journal of the Physical Society of Japan}\ }\textbf
  {\bibinfo {volume} {87}},\ \bibinfo {pages} {074002} (\bibinfo {year}
  {2018})},\ \Eprint
  {http://arxiv.org/abs/https://doi.org/10.7566/JPSJ.87.074002}
  {https://doi.org/10.7566/JPSJ.87.074002} \BibitemShut {NoStop}%
\bibitem [{\citenamefont {Sharir}\ \emph {et~al.}(2020)\citenamefont {Sharir},
  \citenamefont {Levine}, \citenamefont {Wies}, \citenamefont {Carleo},\ and\
  \citenamefont {Shashua}}]{SharirEtAl20}%
  \BibitemOpen
  \bibfield  {author} {\bibinfo {author} {\bibfnamefont {O.}~\bibnamefont
  {Sharir}}, \bibinfo {author} {\bibfnamefont {Y.}~\bibnamefont {Levine}},
  \bibinfo {author} {\bibfnamefont {N.}~\bibnamefont {Wies}}, \bibinfo {author}
  {\bibfnamefont {G.}~\bibnamefont {Carleo}}, \ and\ \bibinfo {author}
  {\bibfnamefont {A.}~\bibnamefont {Shashua}},\ }\href {\doibase
  10.1103/PhysRevLett.124.020503} {\bibfield  {journal} {\bibinfo  {journal}
  {Phys. Rev. Lett.}\ }\textbf {\bibinfo {volume} {124}},\ \bibinfo {pages}
  {020503} (\bibinfo {year} {2020})}\BibitemShut {NoStop}%
\bibitem [{\citenamefont {Torlai}\ and\ \citenamefont
  {Melko}(2018)}]{TorlaiM18}%
  \BibitemOpen
  \bibfield  {author} {\bibinfo {author} {\bibfnamefont {G.}~\bibnamefont
  {Torlai}}\ and\ \bibinfo {author} {\bibfnamefont {R.~G.}\ \bibnamefont
  {Melko}},\ }\href {\doibase 10.1103/PhysRevLett.120.240503} {\bibfield
  {journal} {\bibinfo  {journal} {Phys. Rev. Lett.}\ }\textbf {\bibinfo
  {volume} {120}},\ \bibinfo {pages} {240503} (\bibinfo {year}
  {2018})}\BibitemShut {NoStop}%
\bibitem [{\citenamefont {Glasser}\ \emph {et~al.}(2018)\citenamefont
  {Glasser}, \citenamefont {Pancotti}, \citenamefont {August}, \citenamefont
  {Rodriguez},\ and\ \citenamefont {Cirac}}]{GlasserEtAl18}%
  \BibitemOpen
  \bibfield  {author} {\bibinfo {author} {\bibfnamefont {I.}~\bibnamefont
  {Glasser}}, \bibinfo {author} {\bibfnamefont {N.}~\bibnamefont {Pancotti}},
  \bibinfo {author} {\bibfnamefont {M.}~\bibnamefont {August}}, \bibinfo
  {author} {\bibfnamefont {I.~D.}\ \bibnamefont {Rodriguez}}, \ and\ \bibinfo
  {author} {\bibfnamefont {J.~I.}\ \bibnamefont {Cirac}},\ }\href {\doibase
  10.1103/PhysRevX.8.011006} {\bibfield  {journal} {\bibinfo  {journal} {Phys.
  Rev. X}\ }\textbf {\bibinfo {volume} {8}},\ \bibinfo {pages} {011006}
  (\bibinfo {year} {2018})}\BibitemShut {NoStop}%
\bibitem [{\citenamefont {Liu}\ \emph {et~al.}(2018)\citenamefont {Liu},
  \citenamefont {Zhang}, \citenamefont {Lewenstein},\ and\ \citenamefont
  {Ran}}]{LiuEtAl18}%
  \BibitemOpen
  \bibfield  {author} {\bibinfo {author} {\bibfnamefont {Y.}~\bibnamefont
  {Liu}}, \bibinfo {author} {\bibfnamefont {X.}~\bibnamefont {Zhang}}, \bibinfo
  {author} {\bibfnamefont {M.}~\bibnamefont {Lewenstein}}, \ and\ \bibinfo
  {author} {\bibfnamefont {S.~J.}\ \bibnamefont {Ran}},\ }\href@noop {}
  {\bibfield  {journal} {\bibinfo  {journal} {arXiv preprint arXiv:1803.09111}\
  } (\bibinfo {year} {2018})}\BibitemShut {NoStop}%
\bibitem [{\citenamefont {Pastori}\ \emph {et~al.}(2019)\citenamefont
  {Pastori}, \citenamefont {Kaubruegger},\ and\ \citenamefont
  {Budich}}]{PastoriEtAl19}%
  \BibitemOpen
  \bibfield  {author} {\bibinfo {author} {\bibfnamefont {L.}~\bibnamefont
  {Pastori}}, \bibinfo {author} {\bibfnamefont {R.}~\bibnamefont
  {Kaubruegger}}, \ and\ \bibinfo {author} {\bibfnamefont {J.~C.}\ \bibnamefont
  {Budich}},\ }\href {\doibase 10.1103/PhysRevB.99.165123} {\bibfield
  {journal} {\bibinfo  {journal} {Phys. Rev. B}\ }\textbf {\bibinfo {volume}
  {99}},\ \bibinfo {pages} {165123} (\bibinfo {year} {2019})}\BibitemShut
  {NoStop}%
\bibitem [{\citenamefont {Clark}(2018)}]{Clark18}%
  \BibitemOpen
  \bibfield  {author} {\bibinfo {author} {\bibfnamefont {S.~R.}\ \bibnamefont
  {Clark}},\ }\href@noop {} {\bibfield  {journal} {\bibinfo  {journal} {Journal
  of Physics A: Mathematical and Theoretical}\ }\textbf {\bibinfo {volume}
  {51}},\ \bibinfo {pages} {135301} (\bibinfo {year} {2018})}\BibitemShut
  {NoStop}%
\bibitem [{\citenamefont {Schuld}\ \emph {et~al.}(2014)\citenamefont {Schuld},
  \citenamefont {Sinayskiy},\ and\ \citenamefont
  {Petruccione}}]{Schuld:QInf:2014}%
  \BibitemOpen
  \bibfield  {author} {\bibinfo {author} {\bibfnamefont {M.}~\bibnamefont
  {Schuld}}, \bibinfo {author} {\bibfnamefont {I.}~\bibnamefont {Sinayskiy}}, \
  and\ \bibinfo {author} {\bibfnamefont {F.}~\bibnamefont {Petruccione}},\
  }\href {http://link.springer.com/article/10.1007/s11128-014-0809-8}
  {\bibfield  {journal} {\bibinfo  {journal} {Quantum Information Processing}\
  }\textbf {\bibinfo {volume} {13}},\ \bibinfo {pages} {2567} (\bibinfo {year}
  {2014})}\BibitemShut {NoStop}%
\bibitem [{\citenamefont {Deng}\ \emph {et~al.}(2017)\citenamefont {Deng},
  \citenamefont {Li},\ and\ \citenamefont {Das~Sarma}}]{DengEtAl17}%
  \BibitemOpen
  \bibfield  {author} {\bibinfo {author} {\bibfnamefont {D.~L.}\ \bibnamefont
  {Deng}}, \bibinfo {author} {\bibfnamefont {X.}~\bibnamefont {Li}}, \ and\
  \bibinfo {author} {\bibfnamefont {S.}~\bibnamefont {Das~Sarma}},\ }\href
  {\doibase 10.1103/PhysRevX.7.021021} {\bibfield  {journal} {\bibinfo
  {journal} {Phys. Rev. X}\ }\textbf {\bibinfo {volume} {7}},\ \bibinfo {pages}
  {021021} (\bibinfo {year} {2017})}\BibitemShut {NoStop}%
\bibitem [{\citenamefont {Biamonte}\ \emph {et~al.}(2017)\citenamefont
  {Biamonte}, \citenamefont {Wittek}, \citenamefont {Pancotti}, \citenamefont
  {Rebentrost}, \citenamefont {Wiebe},\ and\ \citenamefont
  {Lloyd}}]{Biamonte:Nat:2017}%
  \BibitemOpen
  \bibfield  {author} {\bibinfo {author} {\bibfnamefont {J.}~\bibnamefont
  {Biamonte}}, \bibinfo {author} {\bibfnamefont {P.}~\bibnamefont {Wittek}},
  \bibinfo {author} {\bibfnamefont {N.}~\bibnamefont {Pancotti}}, \bibinfo
  {author} {\bibfnamefont {P.}~\bibnamefont {Rebentrost}}, \bibinfo {author}
  {\bibfnamefont {N.}~\bibnamefont {Wiebe}}, \ and\ \bibinfo {author}
  {\bibfnamefont {S.}~\bibnamefont {Lloyd}},\ }\href@noop {} {\bibfield
  {journal} {\bibinfo  {journal} {Nature}\ }\textbf {\bibinfo {volume} {549}},\
  \bibinfo {pages} {195} (\bibinfo {year} {2017})}\BibitemShut {NoStop}%
\bibitem [{\citenamefont {Rebentrost}\ \emph {et~al.}(2018)\citenamefont
  {Rebentrost}, \citenamefont {Bromley}, \citenamefont {Weedbrook},\ and\
  \citenamefont {Lloyd}}]{RebentrostEtAl18}%
  \BibitemOpen
  \bibfield  {author} {\bibinfo {author} {\bibfnamefont {P.}~\bibnamefont
  {Rebentrost}}, \bibinfo {author} {\bibfnamefont {T.~R.}\ \bibnamefont
  {Bromley}}, \bibinfo {author} {\bibfnamefont {C.}~\bibnamefont {Weedbrook}},
  \ and\ \bibinfo {author} {\bibfnamefont {S.}~\bibnamefont {Lloyd}},\
  }\href@noop {} {\bibfield  {journal} {\bibinfo  {journal} {Physical Review
  A}\ }\textbf {\bibinfo {volume} {98}},\ \bibinfo {pages} {042308} (\bibinfo
  {year} {2018})}\BibitemShut {NoStop}%
\bibitem [{\citenamefont {Aspuru-Guzik}\ and\ \citenamefont
  {Cao}(2020)}]{AspuruC20}%
  \BibitemOpen
  \bibfield  {author} {\bibinfo {author} {\bibfnamefont {A.}~\bibnamefont
  {Aspuru-Guzik}}\ and\ \bibinfo {author} {\bibfnamefont {Y.}~\bibnamefont
  {Cao}},\ }\href@noop {} {\enquote {\bibinfo {title} {Quantum artificial
  neural networks},}\ } (\bibinfo {year} {2020}),\ \bibinfo {note} {uS Patent
  App. 16/647,194}\BibitemShut {NoStop}%
\bibitem [{\citenamefont {Killoran}\ \emph {et~al.}(2019)\citenamefont
  {Killoran}, \citenamefont {Bromley}, \citenamefont {Arrazola}, \citenamefont
  {Schuld}, \citenamefont {Quesada},\ and\ \citenamefont
  {Lloyd}}]{KilloranEtAl19}%
  \BibitemOpen
  \bibfield  {author} {\bibinfo {author} {\bibfnamefont {N.}~\bibnamefont
  {Killoran}}, \bibinfo {author} {\bibfnamefont {T.~R.}\ \bibnamefont
  {Bromley}}, \bibinfo {author} {\bibfnamefont {J.~M.}\ \bibnamefont
  {Arrazola}}, \bibinfo {author} {\bibfnamefont {M.}~\bibnamefont {Schuld}},
  \bibinfo {author} {\bibfnamefont {N.}~\bibnamefont {Quesada}}, \ and\
  \bibinfo {author} {\bibfnamefont {S.}~\bibnamefont {Lloyd}},\ }\href
  {\doibase 10.1103/PhysRevResearch.1.033063} {\bibfield  {journal} {\bibinfo
  {journal} {Phys. Rev. Research}\ }\textbf {\bibinfo {volume} {1}},\ \bibinfo
  {pages} {033063} (\bibinfo {year} {2019})}\BibitemShut {NoStop}%
\bibitem [{\citenamefont {Mangini}\ \emph {et~al.}(2021)\citenamefont
  {Mangini}, \citenamefont {Tacchino}, \citenamefont {Gerace}, \citenamefont
  {Bajoni},\ and\ \citenamefont {Macchiavello}}]{ManginiEtAl21}%
  \BibitemOpen
  \bibfield  {author} {\bibinfo {author} {\bibfnamefont {S.}~\bibnamefont
  {Mangini}}, \bibinfo {author} {\bibfnamefont {F.}~\bibnamefont {Tacchino}},
  \bibinfo {author} {\bibfnamefont {D.}~\bibnamefont {Gerace}}, \bibinfo
  {author} {\bibfnamefont {D.}~\bibnamefont {Bajoni}}, \ and\ \bibinfo {author}
  {\bibfnamefont {C.}~\bibnamefont {Macchiavello}},\ }\href@noop {} {\bibfield
  {journal} {\bibinfo  {journal} {EPL (Europhysics Letters)}\ }\textbf
  {\bibinfo {volume} {134}},\ \bibinfo {pages} {10002} (\bibinfo {year}
  {2021})}\BibitemShut {NoStop}%
\bibitem [{\citenamefont {Torrontegui}\ and\ \citenamefont
  {Garc{\'\i}a-Ripoll}(2019)}]{TorronteguiG19}%
  \BibitemOpen
  \bibfield  {author} {\bibinfo {author} {\bibfnamefont {E.}~\bibnamefont
  {Torrontegui}}\ and\ \bibinfo {author} {\bibfnamefont {J.~J.}\ \bibnamefont
  {Garc{\'\i}a-Ripoll}},\ }\href@noop {} {\bibfield  {journal} {\bibinfo
  {journal} {EPL (Europhysics Letters)}\ }\textbf {\bibinfo {volume} {125}},\
  \bibinfo {pages} {30004} (\bibinfo {year} {2019})}\BibitemShut {NoStop}%
\bibitem [{\citenamefont {Kristensen}\ \emph {et~al.}(2021)\citenamefont
  {Kristensen}, \citenamefont {Degroote}, \citenamefont {Wittek}, \citenamefont
  {Aspuru-Guzik},\ and\ \citenamefont {Zinner}}]{KristensenEtAl21}%
  \BibitemOpen
  \bibfield  {author} {\bibinfo {author} {\bibfnamefont {L.}~\bibnamefont
  {Kristensen}}, \bibinfo {author} {\bibfnamefont {M.}~\bibnamefont
  {Degroote}}, \bibinfo {author} {\bibfnamefont {P.}~\bibnamefont {Wittek}},
  \bibinfo {author} {\bibfnamefont {A.}~\bibnamefont {Aspuru-Guzik}}, \ and\
  \bibinfo {author} {\bibfnamefont {N.}~\bibnamefont {Zinner}},\ }\href
  {\doibase 10.1038/s41534-021-00381-7} {\bibfield  {journal} {\bibinfo
  {journal} {npj Quantum Information}\ }\textbf {\bibinfo {volume} {7}},\
  \bibinfo {pages} {59} (\bibinfo {year} {2021})}\BibitemShut {NoStop}%
\bibitem [{\citenamefont {Cao}\ \emph {et~al.}(2017)\citenamefont {Cao},
  \citenamefont {Guerreschi},\ and\ \citenamefont {Aspuru-Guzik}}]{CaoGG17}%
  \BibitemOpen
  \bibfield  {author} {\bibinfo {author} {\bibfnamefont {Y.}~\bibnamefont
  {Cao}}, \bibinfo {author} {\bibfnamefont {G.~G.}\ \bibnamefont {Guerreschi}},
  \ and\ \bibinfo {author} {\bibfnamefont {A.}~\bibnamefont {Aspuru-Guzik}},\
  }\href@noop {} {\enquote {\bibinfo {title} {Quantum neuron: an elementary
  building block for machine learning on quantum computers},}\ } (\bibinfo
  {year} {2017}),\ \Eprint {http://arxiv.org/abs/1711.11240} {arXiv:1711.11240
  [quant-ph]} \BibitemShut {NoStop}%
\bibitem [{\citenamefont {Cong}\ \emph {et~al.}(2019)\citenamefont {Cong},
  \citenamefont {Choi},\ and\ \citenamefont {Lukin}}]{CongCL19}%
  \BibitemOpen
  \bibfield  {author} {\bibinfo {author} {\bibfnamefont {I.}~\bibnamefont
  {Cong}}, \bibinfo {author} {\bibfnamefont {S.}~\bibnamefont {Choi}}, \ and\
  \bibinfo {author} {\bibfnamefont {M.~D.}\ \bibnamefont {Lukin}},\ }\href
  {\doibase 10.1038/s41567-019-0648-8} {\bibfield  {journal} {\bibinfo
  {journal} {Nature Physics}\ }\textbf {\bibinfo {volume} {15}},\ \bibinfo
  {pages} {1273–1278} (\bibinfo {year} {2019})}\BibitemShut {NoStop}%
\bibitem [{\citenamefont {Beer}\ \emph {et~al.}(2020)\citenamefont {Beer},
  \citenamefont {Bondarenko}, \citenamefont {Farrelly}, \citenamefont
  {Osborne}, \citenamefont {Salzmann}, \citenamefont {Scheiermann},\ and\
  \citenamefont {Wolf}}]{BeerEtAl20}%
  \BibitemOpen
  \bibfield  {author} {\bibinfo {author} {\bibfnamefont {K.}~\bibnamefont
  {Beer}}, \bibinfo {author} {\bibfnamefont {D.}~\bibnamefont {Bondarenko}},
  \bibinfo {author} {\bibfnamefont {T.}~\bibnamefont {Farrelly}}, \bibinfo
  {author} {\bibfnamefont {T.~J.}\ \bibnamefont {Osborne}}, \bibinfo {author}
  {\bibfnamefont {R.}~\bibnamefont {Salzmann}}, \bibinfo {author}
  {\bibfnamefont {D.}~\bibnamefont {Scheiermann}}, \ and\ \bibinfo {author}
  {\bibfnamefont {R.}~\bibnamefont {Wolf}},\ }\href@noop {} {\bibfield
  {journal} {\bibinfo  {journal} {Nature communications}\ }\textbf {\bibinfo
  {volume} {11}},\ \bibinfo {pages} {1} (\bibinfo {year} {2020})}\BibitemShut
  {NoStop}%
\bibitem [{\citenamefont {Pons}\ \emph {et~al.}(2007)\citenamefont {Pons},
  \citenamefont {Ahufinger}, \citenamefont {Wunderlich}, \citenamefont
  {Sanpera}, \citenamefont {Braungardt}, \citenamefont {Sen(De)}, \citenamefont
  {Sen},\ and\ \citenamefont {Lewenstein}}]{PonsEtAl07}%
  \BibitemOpen
  \bibfield  {author} {\bibinfo {author} {\bibfnamefont {M.}~\bibnamefont
  {Pons}}, \bibinfo {author} {\bibfnamefont {V.}~\bibnamefont {Ahufinger}},
  \bibinfo {author} {\bibfnamefont {C.}~\bibnamefont {Wunderlich}}, \bibinfo
  {author} {\bibfnamefont {A.}~\bibnamefont {Sanpera}}, \bibinfo {author}
  {\bibfnamefont {S.}~\bibnamefont {Braungardt}}, \bibinfo {author}
  {\bibfnamefont {A.}~\bibnamefont {Sen(De)}}, \bibinfo {author} {\bibfnamefont
  {U.}~\bibnamefont {Sen}}, \ and\ \bibinfo {author} {\bibfnamefont
  {M.}~\bibnamefont {Lewenstein}},\ }\href {\doibase
  10.1103/PhysRevLett.98.023003} {\bibfield  {journal} {\bibinfo  {journal}
  {Phys. Rev. Lett.}\ }\textbf {\bibinfo {volume} {98}},\ \bibinfo {pages}
  {023003} (\bibinfo {year} {2007})}\BibitemShut {NoStop}%
\bibitem [{\citenamefont {Gopalakrishnan}\ \emph {et~al.}(2012)\citenamefont
  {Gopalakrishnan}, \citenamefont {Lev},\ and\ \citenamefont
  {Goldbart}}]{GopalakrishnanEtAl12}%
  \BibitemOpen
  \bibfield  {author} {\bibinfo {author} {\bibfnamefont {S.}~\bibnamefont
  {Gopalakrishnan}}, \bibinfo {author} {\bibfnamefont {B.~L.}\ \bibnamefont
  {Lev}}, \ and\ \bibinfo {author} {\bibfnamefont {P.~M.}\ \bibnamefont
  {Goldbart}},\ }\href {\doibase 10.1080/14786435.2011.637980} {\bibfield
  {journal} {\bibinfo  {journal} {Philosophical Magazine}\ }\textbf {\bibinfo
  {volume} {92}},\ \bibinfo {pages} {353} (\bibinfo {year} {2012})},\ \Eprint
  {http://arxiv.org/abs/https://doi.org/10.1080/14786435.2011.637980}
  {https://doi.org/10.1080/14786435.2011.637980} \BibitemShut {NoStop}%
\bibitem [{\citenamefont {Fiorelli}\ \emph {et~al.}(2020)\citenamefont
  {Fiorelli}, \citenamefont {Marcuzzi}, \citenamefont {Rotondo}, \citenamefont
  {Carollo},\ and\ \citenamefont {Lesanovsky}}]{FiorelliEtAl20}%
  \BibitemOpen
  \bibfield  {author} {\bibinfo {author} {\bibfnamefont {E.}~\bibnamefont
  {Fiorelli}}, \bibinfo {author} {\bibfnamefont {M.}~\bibnamefont {Marcuzzi}},
  \bibinfo {author} {\bibfnamefont {P.}~\bibnamefont {Rotondo}}, \bibinfo
  {author} {\bibfnamefont {F.}~\bibnamefont {Carollo}}, \ and\ \bibinfo
  {author} {\bibfnamefont {I.}~\bibnamefont {Lesanovsky}},\ }\href {\doibase
  10.1103/PhysRevLett.125.070604} {\bibfield  {journal} {\bibinfo  {journal}
  {Phys. Rev. Lett.}\ }\textbf {\bibinfo {volume} {125}},\ \bibinfo {pages}
  {070604} (\bibinfo {year} {2020})}\BibitemShut {NoStop}%
\bibitem [{\citenamefont {Behrman}\ \emph {et~al.}(2006)\citenamefont
  {Behrman}, \citenamefont {Gaddam}, \citenamefont {Steck},\ and\ \citenamefont
  {Skinner}}]{Behrman06}%
  \BibitemOpen
  \bibfield  {author} {\bibinfo {author} {\bibfnamefont {E.~C.}\ \bibnamefont
  {Behrman}}, \bibinfo {author} {\bibfnamefont {K.}~\bibnamefont {Gaddam}},
  \bibinfo {author} {\bibfnamefont {J.~E.}\ \bibnamefont {Steck}}, \ and\
  \bibinfo {author} {\bibfnamefont {S.~R.}\ \bibnamefont {Skinner}},\ }\enquote
  {\bibinfo {title} {Microtubules as a quantum hopfield network},}\ in\ \href
  {\doibase 10.1007/3-540-36723-3_10} {\emph {\bibinfo {booktitle} {The
  Emerging Physics of Consciousness}}},\ \bibinfo {editor} {edited by\ \bibinfo
  {editor} {\bibfnamefont {J.~A.}\ \bibnamefont {Tuszynski}}}\ (\bibinfo
  {publisher} {Springer Berlin Heidelberg},\ \bibinfo {address} {Berlin,
  Heidelberg},\ \bibinfo {year} {2006})\ pp.\ \bibinfo {pages}
  {351--370}\BibitemShut {NoStop}%
\bibitem [{\citenamefont {Akazawa}\ \emph {et~al.}(2000)\citenamefont
  {Akazawa}, \citenamefont {Tokuda}, \citenamefont {Asahi},\ and\ \citenamefont
  {Amemiya}}]{AkazawaEtAl00}%
  \BibitemOpen
  \bibfield  {author} {\bibinfo {author} {\bibfnamefont {M.}~\bibnamefont
  {Akazawa}}, \bibinfo {author} {\bibfnamefont {E.}~\bibnamefont {Tokuda}},
  \bibinfo {author} {\bibfnamefont {N.}~\bibnamefont {Asahi}}, \ and\ \bibinfo
  {author} {\bibfnamefont {Y.}~\bibnamefont {Amemiya}},\ }\href@noop {}
  {\bibfield  {journal} {\bibinfo  {journal} {Analog Integrated Circuits and
  Signal Processing}\ }\textbf {\bibinfo {volume} {24}},\ \bibinfo {pages} {51}
  (\bibinfo {year} {2000})}\BibitemShut {NoStop}%
\bibitem [{\citenamefont {Rotondo}\ \emph
  {et~al.}(2018{\natexlab{a}})\citenamefont {Rotondo}, \citenamefont
  {Marcuzzi}, \citenamefont {Garrahan}, \citenamefont {Lesanovsky},\ and\
  \citenamefont {M\"uller}}]{Rotondo:JPA:2018}%
  \BibitemOpen
  \bibfield  {author} {\bibinfo {author} {\bibfnamefont {P.}~\bibnamefont
  {Rotondo}}, \bibinfo {author} {\bibfnamefont {M.}~\bibnamefont {Marcuzzi}},
  \bibinfo {author} {\bibfnamefont {J.~P.}\ \bibnamefont {Garrahan}}, \bibinfo
  {author} {\bibfnamefont {I.}~\bibnamefont {Lesanovsky}}, \ and\ \bibinfo
  {author} {\bibfnamefont {M.}~\bibnamefont {M\"uller}},\ }\href
  {http://stacks.iop.org/1751-8121/51/i=11/a=115301} {\bibfield  {journal}
  {\bibinfo  {journal} {Journal of Physics A: Mathematical and Theoretical}\
  }\textbf {\bibinfo {volume} {51}},\ \bibinfo {pages} {115301} (\bibinfo
  {year} {2018}{\natexlab{a}})}\BibitemShut {NoStop}%
\bibitem [{\citenamefont {Fiorelli}\ \emph {et~al.}(2019)\citenamefont
  {Fiorelli}, \citenamefont {Rotondo}, \citenamefont {Marcuzzi}, \citenamefont
  {Garrahan},\ and\ \citenamefont {Lesanovsky}}]{Fiorelli:PRA:2019}%
  \BibitemOpen
  \bibfield  {author} {\bibinfo {author} {\bibfnamefont {E.}~\bibnamefont
  {Fiorelli}}, \bibinfo {author} {\bibfnamefont {P.}~\bibnamefont {Rotondo}},
  \bibinfo {author} {\bibfnamefont {M.}~\bibnamefont {Marcuzzi}}, \bibinfo
  {author} {\bibfnamefont {J.~P.}\ \bibnamefont {Garrahan}}, \ and\ \bibinfo
  {author} {\bibfnamefont {I.}~\bibnamefont {Lesanovsky}},\ }\href {\doibase
  10.1103/PhysRevA.99.032126} {\bibfield  {journal} {\bibinfo  {journal} {Phys.
  Rev. A}\ }\textbf {\bibinfo {volume} {99}},\ \bibinfo {pages} {032126}
  (\bibinfo {year} {2019})}\BibitemShut {NoStop}%
\bibitem [{\citenamefont {Diamantini}\ and\ \citenamefont
  {Trugenberger}(2006)}]{DiamantiniT06}%
  \BibitemOpen
  \bibfield  {author} {\bibinfo {author} {\bibfnamefont {M.~C.}\ \bibnamefont
  {Diamantini}}\ and\ \bibinfo {author} {\bibfnamefont {C.~A.}\ \bibnamefont
  {Trugenberger}},\ }\href@noop {} {\bibfield  {journal} {\bibinfo  {journal}
  {Phys. Rev. Lett.}\ }\textbf {\bibinfo {volume} {97}},\ \bibinfo {pages}
  {130503} (\bibinfo {year} {2006})}\BibitemShut {NoStop}%
\bibitem [{\citenamefont {Hopfield}(1982)}]{Hopfield:1982}%
  \BibitemOpen
  \bibfield  {author} {\bibinfo {author} {\bibfnamefont {J.~J.}\ \bibnamefont
  {Hopfield}},\ }\href@noop {} {\bibfield  {journal} {\bibinfo  {journal}
  {PNAS}\ }\textbf {\bibinfo {volume} {79}},\ \bibinfo {pages} {2554} (\bibinfo
  {year} {1982})}\BibitemShut {NoStop}%
\bibitem [{\citenamefont {Amit}\ \emph {et~al.}(1985)\citenamefont {Amit},
  \citenamefont {Gutfreund},\ and\ \citenamefont
  {Sompolinsky}}]{Amit:PRL:1985}%
  \BibitemOpen
  \bibfield  {author} {\bibinfo {author} {\bibfnamefont {D.~J.}\ \bibnamefont
  {Amit}}, \bibinfo {author} {\bibfnamefont {H.}~\bibnamefont {Gutfreund}}, \
  and\ \bibinfo {author} {\bibfnamefont {H.}~\bibnamefont {Sompolinsky}},\
  }\href {\doibase 10.1103/PhysRevLett.55.1530} {\bibfield  {journal} {\bibinfo
   {journal} {Phys. Rev. Lett.}\ }\textbf {\bibinfo {volume} {55}},\ \bibinfo
  {pages} {1530} (\bibinfo {year} {1985})}\BibitemShut {NoStop}%
\bibitem [{\citenamefont {Amit}\ \emph {et~al.}(1987)\citenamefont {Amit},
  \citenamefont {Gutfreund},\ and\ \citenamefont {Sompolinsky}}]{AmitGS:1987}%
  \BibitemOpen
  \bibfield  {author} {\bibinfo {author} {\bibfnamefont {D.~J.}\ \bibnamefont
  {Amit}}, \bibinfo {author} {\bibfnamefont {H.}~\bibnamefont {Gutfreund}}, \
  and\ \bibinfo {author} {\bibfnamefont {H.}~\bibnamefont {Sompolinsky}},\
  }\href@noop {} {\bibfield  {journal} {\bibinfo  {journal} {Annals of
  Physics}\ }\textbf {\bibinfo {volume} {173}},\ \bibinfo {pages} {30}
  (\bibinfo {year} {1987})}\BibitemShut {NoStop}%
\bibitem [{\citenamefont {Garrahan}(2018)}]{Garrahan18}%
  \BibitemOpen
  \bibfield  {author} {\bibinfo {author} {\bibfnamefont {J.~P.}\ \bibnamefont
  {Garrahan}},\ }\href {\doibase https://doi.org/10.1016/j.physa.2017.12.149}
  {\bibfield  {journal} {\bibinfo  {journal} {Physica A: Statistical Mechanics
  and its Applications}\ }\textbf {\bibinfo {volume} {504}},\ \bibinfo {pages}
  {130} (\bibinfo {year} {2018})},\ \bibinfo {note} {lecture Notes of the 14th
  International Summer School on Fundamental Problems in Statistical
  Physics}\BibitemShut {NoStop}%
\bibitem [{\citenamefont {Lewenstein}\ \emph {et~al.}(2021)\citenamefont
  {Lewenstein}, \citenamefont {Gratsea}, \citenamefont {Riera-Campeny},
  \citenamefont {Aloy}, \citenamefont {Kasper},\ and\ \citenamefont
  {Sanpera}}]{LewensteinEtAl20}%
  \BibitemOpen
  \bibfield  {author} {\bibinfo {author} {\bibfnamefont {M.}~\bibnamefont
  {Lewenstein}}, \bibinfo {author} {\bibfnamefont {A.}~\bibnamefont {Gratsea}},
  \bibinfo {author} {\bibfnamefont {A.}~\bibnamefont {Riera-Campeny}}, \bibinfo
  {author} {\bibfnamefont {A.}~\bibnamefont {Aloy}}, \bibinfo {author}
  {\bibfnamefont {V.}~\bibnamefont {Kasper}}, \ and\ \bibinfo {author}
  {\bibfnamefont {A.}~\bibnamefont {Sanpera}},\ }\href
  {http://iopscience.iop.org/article/10.1088/2058-9565/ac070f} {\bibfield
  {journal} {\bibinfo  {journal} {Quantum Science and Technology}\ } (\bibinfo
  {year} {2021})}\BibitemShut {NoStop}%
\bibitem [{\citenamefont {Breuer}\ and\ \citenamefont
  {Petruccione}(2002)}]{BreuerP:2002}%
  \BibitemOpen
  \bibfield  {author} {\bibinfo {author} {\bibfnamefont {H.~P.}\ \bibnamefont
  {Breuer}}\ and\ \bibinfo {author} {\bibfnamefont {F.}~\bibnamefont
  {Petruccione}},\ }\href@noop {} {\emph {\bibinfo {title} {The theory of open
  quantum systems}}}\ (\bibinfo  {publisher} {Oxford University Press},\
  \bibinfo {address} {Great Clarendon Street},\ \bibinfo {year}
  {2002})\BibitemShut {NoStop}%
\bibitem [{\citenamefont {M{\"u}ller}\ \emph {et~al.}(2012)\citenamefont
  {M{\"u}ller}, \citenamefont {Diehl}, \citenamefont {Pupillo},\ and\
  \citenamefont {Zoller}}]{MullerEtAl12}%
  \BibitemOpen
  \bibfield  {author} {\bibinfo {author} {\bibfnamefont {M.}~\bibnamefont
  {M{\"u}ller}}, \bibinfo {author} {\bibfnamefont {S.}~\bibnamefont {Diehl}},
  \bibinfo {author} {\bibfnamefont {G.}~\bibnamefont {Pupillo}}, \ and\
  \bibinfo {author} {\bibfnamefont {P.}~\bibnamefont {Zoller}},\ }\href@noop {}
  {\bibfield  {journal} {\bibinfo  {journal} {Advances in Atomic, Molecular,
  and Optical Physics}\ }\textbf {\bibinfo {volume} {61}},\ \bibinfo {pages}
  {1} (\bibinfo {year} {2012})}\BibitemShut {NoStop}%
\bibitem [{\citenamefont {Schindler}\ \emph {et~al.}(2013)\citenamefont
  {Schindler}, \citenamefont {M{\"u}ller}, \citenamefont {Nigg}, \citenamefont
  {Barreiro}, \citenamefont {Martinez}, \citenamefont {Hennrich}, \citenamefont
  {Monz}, \citenamefont {Diehl}, \citenamefont {Zoller},\ and\ \citenamefont
  {Blatt}}]{SchindlerEtAl13}%
  \BibitemOpen
  \bibfield  {author} {\bibinfo {author} {\bibfnamefont {P.}~\bibnamefont
  {Schindler}}, \bibinfo {author} {\bibfnamefont {M.}~\bibnamefont
  {M{\"u}ller}}, \bibinfo {author} {\bibfnamefont {D.}~\bibnamefont {Nigg}},
  \bibinfo {author} {\bibfnamefont {J.~T.}\ \bibnamefont {Barreiro}}, \bibinfo
  {author} {\bibfnamefont {E.~A.}\ \bibnamefont {Martinez}}, \bibinfo {author}
  {\bibfnamefont {M.}~\bibnamefont {Hennrich}}, \bibinfo {author}
  {\bibfnamefont {T.}~\bibnamefont {Monz}}, \bibinfo {author} {\bibfnamefont
  {S.}~\bibnamefont {Diehl}}, \bibinfo {author} {\bibfnamefont
  {P.}~\bibnamefont {Zoller}}, \ and\ \bibinfo {author} {\bibfnamefont
  {R.}~\bibnamefont {Blatt}},\ }\href@noop {} {\bibfield  {journal} {\bibinfo
  {journal} {Nature Physics}\ }\textbf {\bibinfo {volume} {9}},\ \bibinfo
  {pages} {361} (\bibinfo {year} {2013})}\BibitemShut {NoStop}%
\bibitem [{\citenamefont {Kanter}(1988)}]{Kanter88}%
  \BibitemOpen
  \bibfield  {author} {\bibinfo {author} {\bibfnamefont {I.}~\bibnamefont
  {Kanter}},\ }\href {\doibase 10.1103/PhysRevA.37.2739} {\bibfield  {journal}
  {\bibinfo  {journal} {Phys. Rev. A}\ }\textbf {\bibinfo {volume} {37}},\
  \bibinfo {pages} {2739} (\bibinfo {year} {1988})}\BibitemShut {NoStop}%
\bibitem [{\citenamefont {Boll\'e}\ \emph
  {et~al.}(1992{\natexlab{a}})\citenamefont {Boll\'e}, \citenamefont {Dupont},\
  and\ \citenamefont {Huyghebaert}}]{BolleDH_PA_92}%
  \BibitemOpen
  \bibfield  {author} {\bibinfo {author} {\bibfnamefont {D.}~\bibnamefont
  {Boll\'e}}, \bibinfo {author} {\bibfnamefont {P.}~\bibnamefont {Dupont}}, \
  and\ \bibinfo {author} {\bibfnamefont {J.}~\bibnamefont {Huyghebaert}},\
  }\href {\doibase https://doi.org/10.1016/0378-4371(92)90476-7} {\bibfield
  {journal} {\bibinfo  {journal} {Physica A: Statistical Mechanics and its
  Applications}\ }\textbf {\bibinfo {volume} {185}},\ \bibinfo {pages} {363}
  (\bibinfo {year} {1992}{\natexlab{a}})}\BibitemShut {NoStop}%
\bibitem [{\citenamefont {Boll\'e}\ and\ \citenamefont
  {Mallezie}(1989)}]{BolleM_JPA_89}%
  \BibitemOpen
  \bibfield  {author} {\bibinfo {author} {\bibfnamefont {D.}~\bibnamefont
  {Boll\'e}}\ and\ \bibinfo {author} {\bibfnamefont {F.}~\bibnamefont
  {Mallezie}},\ }\href {\doibase 10.1088/0305-4470/22/20/017} {\bibfield
  {journal} {\bibinfo  {journal} {Journal of Physics A: Mathematical and
  General}\ }\textbf {\bibinfo {volume} {22}},\ \bibinfo {pages} {4409}
  (\bibinfo {year} {1989})}\BibitemShut {NoStop}%
\bibitem [{\citenamefont {Boll\'e}\ \emph {et~al.}(1991)\citenamefont
  {Boll\'e}, \citenamefont {Dupont},\ and\ \citenamefont {van
  Mourik}}]{BolleEtAl_JPA_91}%
  \BibitemOpen
  \bibfield  {author} {\bibinfo {author} {\bibfnamefont {D.}~\bibnamefont
  {Boll\'e}}, \bibinfo {author} {\bibfnamefont {P.}~\bibnamefont {Dupont}}, \
  and\ \bibinfo {author} {\bibfnamefont {J.}~\bibnamefont {van Mourik}},\
  }\href {\doibase 10.1088/0305-4470/24/5/021} {\bibfield  {journal} {\bibinfo
  {journal} {Journal of Physics A: Mathematical and General}\ }\textbf
  {\bibinfo {volume} {24}},\ \bibinfo {pages} {1065} (\bibinfo {year}
  {1991})}\BibitemShut {NoStop}%
\bibitem [{\citenamefont {Boll\'e}\ \emph
  {et~al.}(1992{\natexlab{b}})\citenamefont {Boll\'e}, \citenamefont {Dupont},\
  and\ \citenamefont {Huyghebaert}}]{BolleDH_PRA_92}%
  \BibitemOpen
  \bibfield  {author} {\bibinfo {author} {\bibfnamefont {D.}~\bibnamefont
  {Boll\'e}}, \bibinfo {author} {\bibfnamefont {P.}~\bibnamefont {Dupont}}, \
  and\ \bibinfo {author} {\bibfnamefont {J.}~\bibnamefont {Huyghebaert}},\
  }\href {\doibase 10.1103/PhysRevA.45.4194} {\bibfield  {journal} {\bibinfo
  {journal} {Phys. Rev. A}\ }\textbf {\bibinfo {volume} {45}},\ \bibinfo
  {pages} {4194} (\bibinfo {year} {1992}{\natexlab{b}})}\BibitemShut {NoStop}%
\bibitem [{\citenamefont {Wu}(1982)}]{Wu82}%
  \BibitemOpen
  \bibfield  {author} {\bibinfo {author} {\bibfnamefont {F.~Y.}\ \bibnamefont
  {Wu}},\ }\href {\doibase 10.1103/RevModPhys.54.235} {\bibfield  {journal}
  {\bibinfo  {journal} {Rev. Mod. Phys.}\ }\textbf {\bibinfo {volume} {54}},\
  \bibinfo {pages} {235} (\bibinfo {year} {1982})}\BibitemShut {NoStop}%
\bibitem [{\citenamefont {Rotondo}\ \emph
  {et~al.}(2018{\natexlab{b}})\citenamefont {Rotondo}, \citenamefont
  {Marcuzzi}, \citenamefont {Garrahan}, \citenamefont {Lesanovsky},\ and\
  \citenamefont {M\"uller}}]{RotondoEtal:2018}%
  \BibitemOpen
  \bibfield  {author} {\bibinfo {author} {\bibfnamefont {P.}~\bibnamefont
  {Rotondo}}, \bibinfo {author} {\bibfnamefont {M.}~\bibnamefont {Marcuzzi}},
  \bibinfo {author} {\bibfnamefont {J.~P.}\ \bibnamefont {Garrahan}}, \bibinfo
  {author} {\bibfnamefont {I.}~\bibnamefont {Lesanovsky}}, \ and\ \bibinfo
  {author} {\bibfnamefont {M.}~\bibnamefont {M\"uller}},\ }\href
  {http://stacks.iop.org/1751-8121/51/i=11/a=115301} {\bibfield  {journal}
  {\bibinfo  {journal} {Journal of Physics A: Mathematical and Theoretical}\
  }\textbf {\bibinfo {volume} {51}},\ \bibinfo {pages} {115301} (\bibinfo
  {year} {2018}{\natexlab{b}})}\BibitemShut {NoStop}%
\bibitem [{\citenamefont {{Gayrard}}(1992)}]{Gayrard92}%
  \BibitemOpen
  \bibfield  {author} {\bibinfo {author} {\bibfnamefont {V.}~\bibnamefont
  {{Gayrard}}},\ }\href {\doibase 10.1007/BF01048882} {\bibfield  {journal}
  {\bibinfo  {journal} {Journal of Statistical Physics}\ }\textbf {\bibinfo
  {volume} {68}},\ \bibinfo {pages} {977} (\bibinfo {year} {1992})}\BibitemShut
  {NoStop}%
\bibitem [{\citenamefont {Strogatz}(1994)}]{Strogatz94}%
  \BibitemOpen
  \bibfield  {author} {\bibinfo {author} {\bibfnamefont {S.~H.}\ \bibnamefont
  {Strogatz}},\ }\href@noop {} {\emph {\bibinfo {title} {Nonlinear dynamics and
  chaos: with applications to Physics, Biology}}}\ (\bibinfo {year}
  {1994})\BibitemShut {NoStop}%
\bibitem [{\citenamefont {Takeuchi}(1996)}]{Takeuchi96}%
  \BibitemOpen
  \bibfield  {author} {\bibinfo {author} {\bibfnamefont {Y.}~\bibnamefont
  {Takeuchi}},\ }\href@noop {} {\emph {\bibinfo {title} {Global dynamical
  properties of Lotka-Volterra systems}}}\ (\bibinfo  {publisher} {World
  Scientific},\ \bibinfo {year} {1996})\BibitemShut {NoStop}%
\bibitem [{\citenamefont {Goel}\ \emph {et~al.}(1971)\citenamefont {Goel},
  \citenamefont {Maitra},\ and\ \citenamefont {Montroll}}]{NarendraEtAl71}%
  \BibitemOpen
  \bibfield  {author} {\bibinfo {author} {\bibfnamefont {N.~S.}\ \bibnamefont
  {Goel}}, \bibinfo {author} {\bibfnamefont {S.~C.}\ \bibnamefont {Maitra}}, \
  and\ \bibinfo {author} {\bibfnamefont {E.~W.}\ \bibnamefont {Montroll}},\
  }\href {\doibase 10.1103/RevModPhys.43.231} {\bibfield  {journal} {\bibinfo
  {journal} {Rev. Mod. Phys.}\ }\textbf {\bibinfo {volume} {43}},\ \bibinfo
  {pages} {231} (\bibinfo {year} {1971})}\BibitemShut {NoStop}%
\bibitem [{\citenamefont {M{\'e}zard}\ \emph {et~al.}(1990)\citenamefont
  {M{\'e}zard}, \citenamefont {Parisi},\ and\ \citenamefont
  {Virasoro}}]{Mezard:book}%
  \BibitemOpen
  \bibfield  {author} {\bibinfo {author} {\bibfnamefont {M.}~\bibnamefont
  {M{\'e}zard}}, \bibinfo {author} {\bibfnamefont {G.}~\bibnamefont {Parisi}},
  \ and\ \bibinfo {author} {\bibfnamefont {M.-A.}\ \bibnamefont {Virasoro}},\
  }\href@noop {} {\emph {\bibinfo {title} {Spin glass theory and beyond.}}}\
  (\bibinfo  {publisher} {World Scientific Publishing Co., Inc., Pergamon
  Press},\ \bibinfo {year} {1990})\BibitemShut {NoStop}%
\end{thebibliography}%


%
\bibliographystyle{apsrev4-1}

\end{document}